\documentclass[11pt]{article}

\usepackage{amsmath,amssymb,amsthm,mathtools}
\usepackage{authblk}
\usepackage{booktabs,tabularx}
\usepackage{enumitem}
\usepackage{hyperref}

\usepackage[letterpaper,margin=2cm]{geometry}

\newtheorem{theorem}{Theorem}
\newtheorem{lemma}[theorem]{Lemma}
\newtheorem{proposition}[theorem]{Proposition}
\newtheorem{corollary}[theorem]{Corollary}
\theoremstyle{definition}
\newtheorem{definition}[theorem]{Definition}
\newtheorem{example}[theorem]{Example}
\theoremstyle{remark}
\newtheorem{remark}[theorem]{Remark}

\newcommand{\Fp}{\mathbb{F}_p}
\newcommand{\Fps}{\mathbb{F}_p^{\!*}}
\newcommand{\Z}{\mathbb{Z}}
\newcommand{\Zm}{\mathbb{Z}_{p-1}^m}

\newcommand{\CAp}{C(A)^{\perp}}
\newcommand{\dmin}{d_{\min}}
\newcommand{\dQ}{d_Q}
\newcommand{\ellmax}{\ell_{\max}}
\newcommand{\supp}{\mathrm{supp}}
\newcommand{\wt}{\mathrm{wt}}
\newcommand{\CSS}{\mathrm{CSS}}
\newcommand{\QA}{Q(A)}
\newcommand{\DA}{\Delta(A)}
\newcommand{\Qunif}{Q_{\mathrm{unif}}}
\newcommand{\Mult}{\mathrm{Mult}}

\title{Schur saturation and the Clifford ceiling in qudit CSS codes}

\author[1]{Jos\'e Alberto Guzm\'an-Vega}
\author[1]{Juan Carlos Jim\'enez-Cervantes}
\author[1]{Eliseo Sarmiento-Rosales \thanks{Corresponding author: \url{esarmiento@ipn.mx}}}

\affil[1]{Escuela Superior de F\'isica y Matem\'aticas,
Instituto Polit\'ecnico Nacional, 07738 Mexico City, Mexico}

\begin{document}

\maketitle

\begin{abstract}
Orthogonality-type conditions---triorthogonality in the qubit setting
and its prime-dimensional analogues---are the primary algebraic
screening criterion for transversal non-Clifford gate constructions in
CSS codes. In every known positive construction the certificate is
accompanied by an actual non-Clifford logical gate, leaving open
whether a CSS family can satisfy these conditions to arbitrarily high
order while admitting \emph{no} such gate.

We answer this question  for the class of
\emph{symmetric spectral CSS codes}
$Q(A)=\CSS(C(A),C(A))$ built from character evaluation on a finite
abelian group $G$ with $\exp(G)\mid p-1$, $p\geq 5$ prime.
The mechanism is the identity
\[
  C(A)^\perp * C(A)^\perp = \mathbb{F}_p^G,
\]
which we call \emph{Schur saturation} of the dual code.
It follows from the CSS self-orthogonality condition
$A\cap(-A)=\varnothing$ and annihilates every local polynomial phase
of degree~$\geq 3$.
The complete gate-group classification
\[
  S^{\mathrm{CSS}}_{\leq p-1}(A) = C(A)^\perp \oplus C(\mathrm{Stab}(A))
\]
shows that every strictly CSS-preserving local-diagonal transversal
gate lies at level~$\leq 2$ of the Clifford hierarchy---a structural
bound we call the \emph{Clifford ceiling}.
The criterion extends to arbitrary CSS pairs satisfying
$C_Z^\perp * C_Z^\perp = \mathbb{F}_p^n$;
the LDPC instance $[[58,16,3]]_5$ is verified computationally.

The false-positive character of the orthogonality certificates is
sharpest in the Boolean toric family: these codes satisfy
$\ell$-orthogonality for every $\ell\leq p-2$, yet the entire
non-Clifford sector collapses to zero.
The flagship instance $Q_3^*=[[64,56,3]]_5$ encodes $56$~logical
ququints at rate $R=0.875$ with distance-$3$ protection, passes every
orthogonality screening test, and yet admits no local-diagonal
non-Clifford gate of any degree.
The unique non-Pauli survivor is the uniform quadratic drive
$\Qunif$---fault-tolerant, logically nontrivial, and Clifford at
level~$2$, carrying no magic-state resource.
\end{abstract}

 \section{Introduction}\label{sec:intro}
 
\subsection{Context and motivation}
\label{subsec:context}

Fault-tolerant quantum computation requires non-Clifford resources,
often supplied through magic-state distillation~\cite{BK05,BH12,CAB12}.
For CSS codes, the standard route to such gates is to identify
transversal diagonal phases whose existence is suggested by
orthogonality-type conditions:
triorthogonality in the qubit setting~\cite{BH12} and its
prime-dimensional analogues~\cite{CAB12}.
These conditions have acquired the status of \emph{design certificates}:
in every known positive construction, from Bravyi--Haah
triorthogonality~\cite{BH12} to AG-code strong
multiplication~\cite{GG24} and binary Reed--Muller divisibility~\cite{Ngu24}, the
orthogonality-type certificate is accompanied by an actual transversal
non-Clifford logical gate.
This empirical regularity raises a natural question that, to our
knowledge, has not previously been settled:

\begin{center}
\emph{Can a CSS code family satisfy the relevant orthogonality
certificates to arbitrarily high order while admitting no
non-Clifford transversal gate?}
\end{center}

\noindent
A negative answer would confirm that orthogonality certificates
are reliable predictors of non-Clifford gate existence within the
local-diagonal ansatz.
A positive answer---the content of this paper---reveals a structural
gap between algebraic screening and physical gate realization, with
direct consequences for the selection of distillation code blocks.

We construct high-rate qudit CSS codes for which the
orthogonality certificates are satisfied to arbitrarily high order,
yet the complete local-diagonal transversal gate group collapses to
level~$\leq 2$ of the Clifford hierarchy.
The mechanism is not the Eastin--Knill universality
obstruction~\cite{EK09}, but a stronger local-diagonal collapse caused
by \emph{Schur saturation} of the dual code: the identity
\[
  C(A)^\perp * C(A)^\perp = \mathbb{F}_p^G,
\]
which follows directly from the CSS self-orthogonality condition
$A\cap(-A)=\varnothing$, annihilates every local polynomial phase of
degree~$\geq 3$ already under the broad coset-affine relaxation of
Section~\ref{sec:setup}.
The complete gate-group classification arising from this identity---the
\emph{Clifford ceiling}---is derived in closed form and applies
uniformly to symmetric Reed--Solomon, BCH, and Boolean toric CSS codes.

The false-positive character of the orthogonality certificates is
sharpest in the Boolean toric family.
For every prime $p\geq 5$ and every code in this family, the
$\ell$-orthogonality threshold is $\ell_{\max}=p-2$, pointing
algebraically to hierarchy level $p-2$; yet the entire
local-diagonal non-Clifford sector vanishes.
For $p=5$, this is triorthogonality ($\ell=3$), the same certificate that underlies the
Campbell--Anwar--Browne framework~\cite{CAB12}, where additional
algebraic structure promotes it into a transversal cubic-phase gate
$M_\mu=\omega^{\mu\hat{n}^3}$ enabling magic-state distillation.
The flagship instance $Q_3^*=[[64,56,3]]_5$ encodes $56$~logical
ququints at rate $R=0.875$ with distance-$3$ protection, passes
the full triorthogonality screening, and yet admits no such gate;
Table~\ref{tab:physical-profile} summarises this profile.
With logical magic-state distillation recently demonstrated
experimentally on neutral-atom processors~\cite{QuEra25},
structural criteria for selecting distillation code blocks have
acquired direct experimental relevance: Schur saturation provides a
closed-form exclusion test that rules out entire code families from
their spectral support alone, without gate computation or coset
enumeration.

\begin{table}[ht]
\centering
\caption{Screening profile of Full Boolean toric CSS codes.
Rows 1--4 confirm that these codes pass the orthogonality-type
screening tests motivated by known positive constructions;
rows 5--6 show that this evidence does not translate into an actual
local-diagonal non-Clifford gate.
The collapse is driven by Schur saturation of the dual code, not
by Eastin--Knill.
The only non-Pauli transversal survivor is the uniform quadratic
logical Clifford $\Qunif$ (level~$2$), which carries no magic.}
\label{tab:physical-profile}
\begin{tabular}{@{}p{7cm}p{10cm}@{}}
\toprule
\textbf{Property} & \textbf{Status in Full Boolean toric family} \\
\midrule
Distance-$3$ protection &
  \textbf{Yes:} $\dQ=3$ for all codes in the family
  (Theorem~\ref{thm:distance}(a)) \\[3pt]
High rate &
  \textbf{Yes:} $R\to 1$ as $m\to\infty$ for fixed $p$
  (Theorem~\ref{thm:optimal}) \\[3pt]
Triorthogonality ($\ell=3$, $p=5$) &
  \textbf{Yes:} passes triorthogonality screening
  (Theorem~\ref{thm:sharp}) \\[3pt]
$\ell$-orthogonality ($\ell\leq p-2$, general $p$) &
  \textbf{Yes:} passes screening up to level $p-2$
  (Theorem~\ref{thm:sharp}) \\
\midrule
Local-diagonal transversal non-Clifford gate (any degree $\geq 3$) &
  \textbf{No:} Schur saturation collapses every such gate to zero
  (Corollary~\ref{cor:prime-p}) \\[3pt]
Viable CAB-type magic-state distillation block &
  \textbf{No:} screening-level evidence does not imply gate existence
  (Corollary~\ref{cor:prime-p}) \\
\bottomrule
\end{tabular}

\end{table}

 \noindent
The contrast with positive constructions is instructive.
In every known case where an orthogonality certificate accompanies an
actual transversal non-Clifford gate---from Bravyi--Haah
triorthogonality~\cite{BH12} and the Campbell--Anwar--Browne
framework~\cite{CAB12} to AG-code strong
multiplication~\cite{GG24}, binary Reed--Muller
divisibility~\cite{Ngu24}, addressable punctured Reed--Solomon
pairs~\cite{HE25}, and monomial CSS codes~\cite{CLMR26}---the
certificate is accompanied by additional algebraic structure (strong
code multiplication, rich automorphism groups, or decreasing monomial
order) that promotes the screening condition into an actual gate.
The present paper shows that this additional structure is not merely
convenient but \emph{necessary}: this provides, to our knowledge, the first closed-form gate-group
classification of an explicit CSS family exhibiting this false-positive
behavior: $\ell$-orthogonality certificates satisfied to unbounded
order as $p$ grows, yet no non-Clifford gate in the complete
local-diagonal sector.

\subsection{Main contributions}
\label{subsec:contributions}

Section~\ref{sec:rigidity} proves the two core results:
\emph{Schur saturation} (Theorem~\ref{thm:schur}), showing
$C(A)^\perp * C(A)^\perp = \mathbb{F}_p^G$ for every $A$ with
$A\cap(-A)=\varnothing$; and the \emph{Clifford ceiling}
(Theorem~\ref{thm:clifford-ceiling}), giving the complete
local-diagonal gate classification
$S^{\mathrm{CSS}}_{\leq p-1}(A) = C(A)^\perp \oplus C(\mathrm{Stab}(A))$,
entirely at level~$\leq 2$ of the Clifford hierarchy.
Both extend to arbitrary CSS pairs
(Theorem~\ref{thm:general-schur-obstruction}); an exact
computationally checkable criterion is given in
Lemma~\ref{lem:schur-criterion}.
Section~\ref{sec:applications} applies the ceiling to symmetric
Reed--Solomon, BCH, and Boolean toric codes, establishes the sharp
$\ell$-orthogonality threshold $\ell_{\max}=p-2$
(Theorem~\ref{thm:sharp}, Corollary~\ref{cor:prime-p}),
and verifies the ceiling for the LDPC instance
$[[58,16,3]]_5$ (Corollary~\ref{cor:hgp-58-16-3}).

\section{Physical consequences and fault-tolerance interpretation}
\label{sec:physics}

This section presents the physical interpretation and implications
of the main results, with an emphasis on fault-tolerant quantum
computation.
Formal statements and proofs are deferred to
Sections~\ref{sec:setup}--\ref{sec:applications}.
Readers interested primarily in the algebraic framework may proceed
directly to Section~\ref{sec:setup} and return here after
Corollary~\ref{cor:prime-p}.
 
\subsection{The Clifford hierarchy and local diagonal gates}
\label{subsec:hierarchy}
 
For a single qudit of prime dimension~$p$, a diagonal gate
$U_f=\sum_{x\in\mathbb{F}_p}\omega^{f(x)}\lvert x\rangle\langle x\rvert$,
$\omega=e^{2\pi i/p}$, sits at level~$\deg f$ of the Clifford hierarchy,
where $\deg f$ is the degree of the unique polynomial representative of
$f:\mathbb{F}_p\to\mathbb{F}_p$ of degree~$\leq p-1$.
This is the content of Theorem~2 of Cui, Gottesman, and
Krishna~\cite{CGK17}, specialised to precision~$m=1$.
For $p=5$ ququints the dictionary is:
 
\begin{center}
\renewcommand{\arraystretch}{1.25}
\begin{tabular}{@{}ccl@{}}
\toprule
Degree of $f$ & Hierarchy level & Physical gate \\
\midrule
$1$ & $1$ & Pauli $Z$ \\
$2$ & $2$ & Diagonal Clifford (quadratic phase $D_\beta$) \\
$3$ & $3$ & Non-Clifford: cubic phase
              $M_\mu=\omega^{\mu\hat{n}^3}$ \\
$4$ & $4$ & Fourth-level non-Clifford \\
\bottomrule
\end{tabular}
\end{center}
 
\noindent
The same dictionary holds for every prime $p\geq 5$, with degrees running
from $1$ to $p-1$.
The orthogonality threshold $\ellmax=p-2$, established in
Section~\ref{sec:applications}, certifies that the Boolean toric
regime passes the screening test for a transversal diagonal gate
candidate at level $p-2$ of the Clifford hierarchy.
The main result shows that no such gate exists.
 
\subsection{The certificate and its failure: a worked example on
$Q_3^*=[[64,56,3]]_5$}
\label{subsec:worked-example}

The following argument traces, for $Q_3^*=[[64,56,3]]_5$,
how the Schur-saturation collapse operates when one attempts
to implement the transversal cubic-phase gate
\[
  M_\mu
  = \sum_{x\in\mathbb{F}_5}\omega^{\mu x^3}\lvert x\rangle\langle x\rvert
  = \omega^{\mu\hat{n}^3},
\]
the natural non-Clifford target in the ququint setting
used by Campbell, Anwar, and Browne~\cite{CAB12} for
magic-state distillation.
Since $Q_3^*$ satisfies triorthogonality
(Theorem~\ref{thm:sharp}), it passes the same screening condition that,
in CAB-type positive constructions with additional algebraic structure,
leads to a logical non-Clifford gate: $M_\mu^{\otimes 64}$ would
implement such an operation if the certificate were sufficient.
Full proofs appear in Section~\ref{sec:rigidity}.

\medskip
\noindent\textbf{Step 1: the compatibility check.}
To implement $M_\mu^{\otimes 64}$ transversally, a necessary
algebraic test is that the induced phase
$F(x)=\mu\sum_{P\in G}x_P^3$ vary affinely along cosets of
$C(A)$ inside $C(A)^\perp$ (the CASI criterion,
Definition~\ref{def:casi} in Section~\ref{sec:setup}):
$F(r+c)-F(r)$ must be independent of $r\in C(A)^\perp$
for every $c\in C(A)$.

\medskip
\noindent\textbf{Step 2: the Schur-saturation collapse.}
The algebraic analysis of Section~\ref{sec:rigidity} shows that
this check fails for \emph{every} nonzero cubic phase.
Expanding $H_c(r)=F(r+c)-F(r)-F(c)+F(0)$, separating by
degree in~$r$, and polarising, the cubic coefficient
$\alpha$ must satisfy
\[
  \alpha * c\;\in\;
  \bigl(C(A)^\perp * C(A)^\perp\bigr)^\perp = \{0\},
\]
where the last equality is Schur saturation of the dual code
(Section~\ref{sec:rigidity}).
Since $c=\chi_{\alpha_0}$ for any $\alpha_0\in A$ is a
character on the torus with all coordinates nonzero,
$\alpha*c=0$ forces $\alpha=0$.
The cubic phase is annihilated entirely; the argument extends to
every degree~$\geq 3$ by induction on the leading degree.

\medskip
\noindent\textbf{Step 3: what survives.}
Section~\ref{sec:applications} gives the complete surviving sector.
The only strictly CSS-preserving local diagonal phases on
$Q_3^*$ are logical $Z$-type linear phases (parametrised
by $\gamma\in C(A)^\perp$) and the uniform quadratic drive
$\Qunif=\sum_{P\in G}x_P^2$.
By the Cui--Gottesman--Krishna correspondence~\cite{CGK17},
$\Qunif$ sits at level~$2$ of the Clifford hierarchy and
carries no magic.
For prime $p$, Lagrange interpolation makes this classification
exhaustive over all local diagonal gates in the computational basis.
 
\subsection{Physical status of the surviving gate}
\label{subsec:qunif-status}
 
The quadratic survivor $\Qunif$ deserves a careful physical reading.
 
\noindent\textbf{Logically nontrivial.}
As shown in Section~\ref{sec:applications}, $\Qunif$ descends to a
well-defined quadratic form $q_{\mathrm{unif}}([r])=\sum_{P\in G}r_P^2$
on the logical space $C(A)^\perp/C(A)$, whose polar form
$B_{\mathrm{unif}}([r],[s])=2\langle r,s\rangle$ is nondegenerate.
Hence $\Qunif$ induces a nontrivial logical operation; for $Q_3^*$, the
induced logical quadratic form has rank~$56$.
 
\medskip
\noindent\textbf{Fault-tolerant.}
Since $\Qunif$ is implemented as a product of single-qudit quadratic
phase gates applied independently to each of the $n=64$ physical
qudits, it is transversal in the standard sense.
A single physical fault on one qudit cannot propagate to others within the
block.
Combined with the distance-$3$ protection of $Q_3^*$, the code can detect
and correct any single-qudit error occurring before or after the gate.
 
\medskip
\noindent\textbf{Clifford, not magic.}
Despite being logically nontrivial and fault-tolerant, $\Qunif$ is a
diagonal Clifford gate at level~$2$.
It carries no magic-state resource, requires no distillation overhead, and
cannot on its own generate non-Clifford logical resources.
 
\subsection{The no-go for Campbell--Anwar--Browne-type distillation}
\label{subsec:nogo-cab}
 
The Campbell--Anwar--Browne protocols~\cite{CAB12} for prime-dimensional
qudits require a code block equipped with a transversal gate at
level~$\geq 3$ of the Clifford hierarchy.
 
\begin{theorem}[No-go for CAB-type distillation; proved in
Section~\ref{sec:applications}]
\label{thm:nogo-informal}
For every prime $p\geq 5$ and every Full Boolean toric CSS code
$\QA$ with $\dQ=3$, no strictly CSS-preserving local diagonal gate
$F(x)=\sum_{P\in G}f_P(x_P)$ implements a logical non-Clifford operation.
The complete surviving sector $S^{\mathrm{CSS}}_{\leq p-1}(A)=
C(A)^\perp\oplus\langle\Qunif\rangle$ lies entirely at level~$\leq 2$
of the Clifford hierarchy; in particular, no code in this family can
serve as a CAB-type distillation block within the local-diagonal ansatz.
\end{theorem}

\noindent
The complete proof appears in
Sections~\ref{sec:rigidity} and~\ref{sec:applications}.
 
This conclusion holds for \emph{all} prime $p\geq 5$:
for $p=7$ the orthogonality threshold reaches $\ellmax=5$, pointing
algebraically to hierarchy level~$5$, yet the complete local-diagonal
sector collapses to level~$\leq 2$.

\subsection{Relation to the Eastin--Knill theorem}
\label{subsec:eastin-knill}
 
The Eastin--Knill theorem~\cite{EK09} shows that no quantum
error-correcting code detecting arbitrary single-site errors can have a
universal set of transversal gates, ruling out universality but leaving open
which specific non-Clifford gates are achievable.
Our result occupies a different logical position:
\begin{itemize}[leftmargin=1.5em]
\item \emph{Eastin--Knill:} no universal transversal gate set exists.
\item \emph{Clifford ceiling:} in the symmetric spectral CSS regime, not
  even a single local-diagonal non-Clifford gate in the computational basis
  exists within the single-site polynomial ansatz, despite $\ell$-orthogonality
  being satisfied to arbitrarily high order as $p$ grows.
\end{itemize}
The mechanisms are also different.
Eastin--Knill invokes the incompatibility between error detection and
universality.
The Clifford ceiling derives from the identity
$C(A)^\perp*C(A)^\perp=\mathbb{F}_p^G$, which makes the dual code so large
that every CASI local-polynomial phase of degree~$\geq 3$ must vanish
identically, with no reference to error-correction properties.
The dual code is not too \emph{small} to support non-Clifford action; it is
too \emph{large} to permit it.

\section{Spectral CSS Codes}\label{sec:setup}

\subsection{Spectral codes and the CSS condition}

Let \(p\geq 5\) be a prime, \(m\geq 2\), and
\(G=(\mathbb{F}_p^*)^m\) the \(m\)-dimensional algebraic torus over
\(\mathbb{F}_p\). Let $g$ be a generator of $\mathbb{F}_p^*$, and let $e_1, \dots, e_m$ be the standard generators of $G$. Any $x \in G$ decomposes uniquely as $x = e_1^{k_1}\cdots e_m^{k_m}$; for each coordinate, we denote $x_j \coloneq g^{k_j} \in \mathbb{F}_p^*$. We identify the character group $\widehat{G}\cong \mathbb{Z}_{p-1}^m$ via $\alpha\mapsto\chi_\alpha$, where $\chi_\alpha(x)=x_1^{\alpha_1}\cdots x_m^{\alpha_m}$. These characters satisfy a standard orthogonality relation over $G$: $\sum_{x \in G} \chi_\alpha(x)\chi_\beta(x) = |G| \delta_{\alpha, -\beta}$.

\begin{definition}[Spectral code]
\label{def:spectral}
For $A\subseteq \mathbb{Z}_{p-1}^m$, the \emph{spectral code} is
$$
  C(A) \;:=\;
  \text{span}_{\mathbb{F}_p}\{\chi_\alpha:\alpha\in A\}
  \;\subseteq\;
  \mathbb{F}_p^G.
$$
When $C(A)\subseteq C(A)^{\perp}$, the \emph{symmetric CSS code}~\cite{CS96,Ste96} attached to $A$ is $Q(A):=\text{CSS}(C(A),C(A))$. By character orthogonality, we will prove that the inner product of two codewords vanishes unless their spectral indices sum to zero; thus $C(A) \subseteq C(A)^\perp$ if and only if $A \cap (-A) = \emptyset$.
\end{definition}

\begin{proposition}[Parameters]
\label{prop:params}
$\dim C(A)=|A|$. When well-defined, $\mathcal{Q}(A)$ has parameters $[[(p-1)^m,\,(p-1)^m-2|A|,\,d_Q]]_p$~\cite{KLKS06}.
\end{proposition}

\begin{proof}
Linear independence of characters~\cite{Ter99} gives $\dim C(A)=|A|$.
The logical dimension follows from $k=\dim C(A)^\perp - \dim C(A) = 
(n-|A|)-|A|$, where $n=|G|=(p-1)^m$.
\end{proof}

\begin{definition}[Sumsets and negation]
\label{def:sumsets}
For $A\subseteq \mathbb{Z}_{p-1}^m$ and $\ell\geq 1$, set
$\ell A:=\{a_1+\cdots+a_\ell:a_i\in A\}$
and $-A:=\{-\alpha:\alpha\in A\}$, with arithmetic performed modulo $p-1$ coordinate-wise.
\end{definition}

\subsection{$\ell$-orthogonality}

\begin{definition}[$\ell$-orthogonality]
\label{def:l-ortho}
A linear code $C\subseteq\mathbb{F}_p^G$ is \emph{$\ell$-orthogonal}
for an integer $\ell\geq 2$ if every coordinate-wise product of
$\ell-1$ codewords is orthogonal to $C$:
\[
  \langle f_1\cdots f_{\ell-1},\,g\rangle = 0
  \qquad\text{for all }f_1,\dots,f_{\ell-1},\,g\in C.
\]
\end{definition}

\begin{remark}[Triorthogonality, convention, and Clifford hierarchy]
\label{rem:triorthogonality}
For $\ell=3$, Definition~\ref{def:l-ortho} recovers the
\emph{triorthogonality} of Bravyi--Haah~\cite{BH12}: the condition
$\langle f_1 f_2, g\rangle=0$ involves three codewords, consistent with
the prefix ``tri.''
This convention agrees with Rengaswamy et~al.~\cite{RCKP20}, who define
triorthogonality as $\langle f_1 f_2, f_3\rangle=0$ for all
$f_1,f_2,f_3\in C$; our $\ell=3$ reads $\langle f_1 f_2, g\rangle=0$
for all $f_1,f_2,g\in C$, which is the same condition.
An alternative convention found in some literature counts the total number
of codewords in the product on the left-hand side; under that convention
our $\ell$ would appear as $\ell-1$.
We follow the Bravyi--Haah--Rengaswamy convention throughout: an
$\ell$-orthogonal code passes the orthogonality-type screening test
motivated by known positive constructions for a transversal diagonal gate
candidate at level~$\ell$ of the Clifford hierarchy.
 
\end{remark}

\begin{theorem}[Spectral $\ell$-orthogonality criterion]
\label{thm:criterion}
$C(A)$ is $\ell$-orthogonal, that is,
$\langle f_1\cdots f_{\ell-1},\, g\rangle=0$ for all
$f_1,\dots,f_{\ell-1},\, g\in C(A)$, if and only if
$((\ell-1)A)\cap(-A)=\emptyset$ in $\mathbb{Z}_{p-1}^m$.
\end{theorem}

\begin{proof}
Parseval's identity on $G$ relates the spatial inner product to
the spectral domain:
\[\langle f, g \rangle = |G|^{-1}\sum_{\psi}\hat{f}(\psi)\hat{g}(-\psi).\]
Let $f = f_1 \cdots f_{\ell-1}$.
In Fourier analysis, the transform of a pointwise product is the
convolution of their individual transforms $\hat{f_i}$.
Because each $f_i \in C(A)$, their spectral support is restricted
to $\mathrm{supp}(\hat{f}_i) \subseteq A$.
The convolution rule dictates that the support of the product is
contained within the Minkowski sum of the individual supports:
$\mathrm{supp}(\widehat{f_1 \cdots f_{\ell-1}}) \subseteq (\ell-1)A$.
For the inner product sum to be non-zero, there must exist at least
one frequency $\psi$ where both $\hat{f}(\psi) \neq 0$ and
$\hat{g}(-\psi) \neq 0$, requiring $\psi \in (\ell-1)A$ and
$\psi \in -A$.
Therefore, non-vanishing demands
$\psi \in ((\ell-1)A) \cap (-A)$.
The converse is constructed by taking $f_i = \chi_{a_i}$ with
$a_1 + \cdots + a_{\ell-1} = \psi$,
and $g = \chi_{-\psi}$.
\end{proof}

\begin{corollary}[Self-orthogonality]
\label{cor:selforth}
$C(A)\subseteq C(A)^{\perp}$ if and only if $A\cap(-A)=\emptyset$.
\end{corollary}

\begin{proof}
Apply Theorem~\ref{thm:criterion} with $\ell=2$.
\end{proof}

\subsection{The CASI criterion}

\begin{definition}[CASI criterion and strict CSS preservation]
\label{def:casi}
Let \(C=C(A)\). A local diagonal phase \(F:\mathbb{F}_p^G\to\mathbb{F}_p\) satisfies the \emph{CASI criterion} (Coset-Affine Shift Independence) if, for every \(c\in C\), the shift
\begin{equation}
\label{eq:casi}
F(r+c)-F(r)
\end{equation}
is independent of \(r\in C^{\perp}\). Equivalently, there exists a function \(\phi:C\to\mathbb{F}_p\) such that
\(
F(r+c)=F(r)+\phi(c), r\in C^{\perp},\; c\in C.
\)
We say that \(F\) is \emph{strictly CSS-preserving} if
\(
F(r+c)=F(r), r\in C^{\perp},\; c\in C.
\)
Equivalently, \(F\) satisfies the CASI criterion with \(\phi=0\).
\end{definition}

\begin{lemma}[Linearity of the CASI shift functional]
\label{lem:casi-linear}
If $F$ satisfies the CASI criterion with function $\phi:C\to\mathbb{F}_p$,
then $\phi$ is $\mathbb{F}_p$-linear on $C$.
\end{lemma}

\begin{proof}
For $c,d\in C$ and any $r\in C^\perp$, applying the CASI identity twice
and using $C\subseteq C^\perp$ (so $r+c\in C^\perp$ when $r\in C^\perp$):
\begin{align*}
    \phi(c+d)&
    = F(r+c+d)-F(r)\\&
  = [F(r+c+d)-F(r+c)] + [F(r+c)-F(r)]\\&
  = \phi(d)+\phi(c). \qedhere
\end{align*}
\end{proof}

\begin{remark}\label{rem:casi}
The shift $F(r+c)-F(r)$ arises naturally when a diagonal gate
$U_F = \sum_x \omega^{F(x)}\lvert x\rangle\langle x\rvert$ acts on a
CSS coset state
$\lvert r+C\rangle = \lvert C\rvert^{-1/2}\sum_{c\in C}\lvert r+c\rangle$:
the gate produces
$\omega^{F(r)}\lvert C\rvert^{-1/2}\sum_{c\in C}
\omega^{F(r+c)-F(r)}\lvert r+c\rangle$,
so the internal phase pattern is exactly the shift.
The CASI criterion requires this shift to be independent of the coset
representative~$r$, meaning $U_F$ introduces the same internal twist
$\phi(c)$ on every logical coset.
By Lemma~\ref{lem:casi-linear}, $\phi$ is $\mathbb{F}_p$-linear on $C$.
Hence CASI is equivalent to strict CSS preservation modulo a fixed
Pauli-$Z$ correction.
Actual preservation of the CSS codespace requires the stronger strict
condition $\phi=0$.
The two results of Theorem~\ref{thm:clifford-ceiling} and
Corollary~\ref{cor:boolean-ceiling} are therefore read as follows:
\begin{itemize}[leftmargin=1.5em]
\item \emph{Algebraic (CASI):} even under this broad relaxation, every
local phase of degree $\geq 3$ vanishes.
\item \emph{Physical (strict CSS):} the codespace-preserving sector is
$C(A)^{\perp}\oplus C(\mathrm{Stab}(A))$, entirely Clifford.
\end{itemize}
\end{remark}

\section{Schur Saturation and Clifford Rigidity}\label{sec:rigidity}

This section is the algebraic core of the paper.
Sections~\ref{subsec:schur-sat}--\ref{subsec:classification} prove
Schur saturation and the Clifford ceiling for symmetric spectral CSS
codes.
Section~\ref{subsec:general-schur-obstruction} extends the mechanism
to arbitrary CSS pairs, isolating the two operative
conditions---Schur saturation of $C_Z^\perp$ and full coordinate
support of $C_X$---that drive the ceiling in any code-family setting.

\subsection{Schur saturation of the dual code}\label{subsec:schur-sat}

\begin{lemma}[Spectral description of the dual code]
\label{lem:dual-spectral}
Let $G$ be a finite abelian group with $\exp(G)\mid p-1$,
$H=\widehat{G}$, and $B:=H\setminus(-A)$. Then
$C(A)^\perp = C(B)$.
\end{lemma}

\begin{proof}
By character orthogonality, $\langle\chi_\alpha,\chi_\beta\rangle = |G|\,\delta_{\alpha+\beta,0}$
in $\mathbb{F}_p$. Hence $\chi_\beta\perp C(A)$ if and only if
$\beta\notin -A$, i.e., $\beta\in B$. Since $\{\chi_\beta:\beta\in H\}$
is a basis of $\mathbb{F}_p^G$, the conclusion follows.
\end{proof}

\begin{lemma}[Majority-sum lemma]
\label{lem:majority}
Let $H$ be a finite abelian group and $B\subseteq H$ with $|B|>|H|/2$.
Then $B+B=H$.
\end{lemma}

\begin{proof}
For any $\eta\in H$, the sets $B$ and $\eta-B$ satisfy
$|B|+|\eta-B|=2|B|>|H|$, so they cannot be disjoint. Any
$b\in B\cap(\eta-B)$ satisfies $\eta=b+b'$ for some $b'\in B$.
\end{proof}

\begin{theorem}[Schur saturation]
\label{thm:schur}
Let $p\geq 5$ be prime,\footnote{The hypothesis $p\geq 5$
is sharp: for $p=3$, negation is the identity in
$\mathbb{Z}_2^m$, so $A\cap(-A)=A\neq\varnothing$ for
every nonempty $A$; and the polarization uses $d^*!\neq 0$
in $\mathbb{F}_p$, which fails for $d^*\geq 3$ in
$\mathbb{F}_3$.}
$G$ a finite abelian group with $\exp(G)\mid p-1$,
$H=\widehat{G}$, and $A\subseteq H$ nonempty with $A\cap(-A)=\varnothing$.
Then
\[
  C(A)^\perp * C(A)^\perp = \mathbb{F}_p^G.
\]
In particular, this holds for $G=(\mathbb{F}_p^*)^m$, $m\geq 2$, and
any $A\subseteq\{0,1\}^m\setminus\{0\}$.
\end{theorem}

\begin{proof}
Set $B:=H\setminus(-A)$. By Lemma~\ref{lem:dual-spectral},
$C(A)^\perp=C(B)$. The pointwise product of characters satisfies
$\chi_\beta*\chi_\gamma=\chi_{\beta+\gamma}$, so
\(
  C(A)^\perp*C(A)^\perp
  = C(B)*C(B)
  = \mathrm{span}\{\chi_{\beta+\gamma}:\beta,\gamma\in B\}
  = C(B+B).
\)
It suffices to show $B+B=H$. The condition $A\cap(-A)=\varnothing$
forces $A\cap H[2]=\varnothing$, where $H[2]:=\{\alpha\in H:2\alpha=0\}$,
since every $\alpha\in H[2]$ satisfies $-\alpha=\alpha$. Elements of
$H\setminus H[2]$ pair as $\{\alpha,-\alpha\}$ with $\alpha\neq -\alpha$,
and $A$ contains at most one from each pair. Therefore
\(
  |A|\leq\frac{|H|-|H[2]|}{2},
\)
giving $|B|=|H|-|A|\geq(|H|+|H[2]|)/2>|H|/2$. By the
majority-sum lemma (Lemma~\ref{lem:majority}), $B+B=H$.
\end{proof}

\begin{corollary}[Iterated Schur saturation]
\label{cor:iter-schur}
Under the hypotheses of Theorem~\ref{thm:schur}, for every $k\geq 2$,
\[
  \bigl(C(A)^\perp\bigr)^{*k} = \mathbb{F}_p^G.
\]
\end{corollary}

\begin{proof}
The base case $k=2$ is Theorem~\ref{thm:schur}.
For $k\geq 3$: since $C(A)^\perp*C(A)^\perp=\mathbb{F}_p^G$, the
subspace $C(A)^\perp$ has full coordinate support on $G$---if some
$P\in G$ were zero for every $u\in C(A)^\perp$, then $\delta_P$ would
not lie in $C(A)^\perp*C(A)^\perp=\mathbb{F}_p^G$, a contradiction.
Full coordinate support gives $\mathbb{F}_p^G*C(A)^\perp=\mathbb{F}_p^G$:
for each $P$, pick $u\in C(A)^\perp$ with $u_P\neq 0$; then
$u_P^{-1}\delta_P*u=\delta_P\in\mathbb{F}_p^G*C(A)^\perp$.
The inductive step follows:
$(C(A)^\perp)^{*k}=(C(A)^\perp)^{*(k-1)}*C(A)^\perp
\supseteq\mathbb{F}_p^G*C(A)^\perp=\mathbb{F}_p^G$.
\end{proof}

\subsection{Gate collapse: degree $\geq 3$ vanishes}\label{subsec:gate-collapse}

\begin{theorem}[Cubic no-go / Absence of transversal $M_\mu$]
\label{thm:cubic}
Let \(p\geq 5\) be prime, $G$ a finite abelian group with
$\exp(G)\mid p-1$, $H=\widehat{G}$, and $A\subseteq H$
nonempty with $A\cap(-A)=\varnothing$. If
\[
  F(x)=\sum_{P\in G}\bigl(\alpha_P x_P^3+\beta_P x_P^2+\gamma_P x_P\bigr)
\]
satisfies the CASI criterion~\eqref{eq:casi}, then \(\alpha=0\).
\end{theorem}

\begin{proof}
Fix \(c\in C(A)\) and define
\( 
  H_c(r)\;:=\;F(r+c)-F(r)-F(c)+F(0).
\)
By~\eqref{eq:casi}, \(F(r+c)-F(r)\) depends only on \(c\); evaluating
at \(r=0\) gives \(F(r+c)-F(r)=F(c)-F(0)\), so \(H_c(r)\equiv 0\) on
\(C(A)^{\perp}\). Expanding using
\((r+c)^3-r^3=3r^2c+3rc^2+c^3\) and \((r+c)^2-r^2=2rc+c^2\),
\[
  H_c(r)=\underbrace{\sum_P 3\alpha_P c_P\,r_P^2}_{=:Q_c(r)}
        +\underbrace{\sum_P (3\alpha_P c_P^2+2\beta_P c_P)\,r_P}_{=:L_c(r)}.
\]

\emph{Separation by degree in \(r\).} Since \(C(A)^{\perp}\) is an
\(\Fp\)-subspace, \(\lambda r\in C(A)^{\perp}\) for every
\(\lambda\in\Fp\). Substituting:
\( 
  H_c(\lambda r)=\lambda^2 Q_c(r)+\lambda L_c(r)=0
  \ \forall\lambda\in\Fp.
\)
Choosing three distinct \(\lambda_1,\lambda_2,\lambda_3\in\Fp\)
(possible since \(|\Fp|\geq 5\)), the resulting Vandermonde system
forces \(Q_c(r)=0\) and \(L_c(r)=0\) separately for every
\(r\in C(A)^{\perp}\).

\emph{Polarization.} Since \(p\geq 5\), we have \(2\neq 0\) in \(\mathbb{F}_p\),
so the bilinear form associated with \(Q_c\) is recoverable:
\[
  Q_c(r+s)-Q_c(r)-Q_c(s)=6\sum_P\alpha_P c_P r_P s_P=0
  \quad\forall r,s\in C(A)^{\perp}.
\]
Since \(p\geq 5\), we have \(6\neq 0\) in \(\mathbb{F}_p\). The identity
asserts that
\(\alpha*c\) is orthogonal to every coordinate-wise product
\(r*s\) with \(r,s\in C(A)^{\perp}\), i.e.,
\(  \alpha*c\;\in\;\bigl(C(A)^{\perp}*C(A)^{\perp}\bigr)^{\perp}.
\)

\emph{Schur saturation closes the argument.} By
Theorem~\ref{thm:schur}, \(C(A)^{\perp}*C(A)^{\perp}=\Fp^G\), whose
orthogonal complement is \(\{0\}\). Hence \(\alpha*c=0\) for every
\(c\in C(A)\). Pick any \(\alpha_0\in A\); then \(c=\chi_{\alpha_0}\in
C(A)\) has all coordinates nonzero (it is a character on the torus),
and \(\alpha*c=0\) forces \(\alpha=0\).
\end{proof}

\begin{theorem}[Degree-\(d\) no-go]
\label{thm:degree-d-nogo}
Let \(p\geq 5\) be prime, $G$ a finite abelian group with
$\exp(G)\mid p-1$, $H=\widehat{G}$, $A\subseteq H$ nonempty
with $A\cap(-A)=\varnothing$, and \(2\leq d\leq p-1\). If
\[
  F(x)=\sum_{P\in G}\sum_{k=1}^{d}\alpha_P^{(k)}\,x_P^k
\]
satisfies the CASI criterion~\eqref{eq:casi}, then
\(\alpha_P^{(k)}=0\) for every \(k\geq 3\) and every \(P\in G\).
\end{theorem}

\begin{proof}
We argue by descent on the highest degree
\(d^{*}=\max\{k:\alpha^{(k)}\neq 0\}\). Suppose
\(d^{*}\geq 3\). The component of \(H_c(r)=F(r+c)-F(r)-F(c)+F(0)\)
of degree \(d^{*}-1\) in \(r\) is
\[
  d^{*}\cdot \sum_P \alpha_P^{(d^{*})}\,c_P\,r_P^{d^{*}-1},
\]
since \((r+c)^{d^{*}}=\sum_{j}\binom{d^{*}}{j}r^j c^{d^{*}-j}\) and
the contribution of degree \(d^{*}-1\) in \(r\) has binomial coefficient
\(\binom{d^{*}}{d^{*}-1}=d^{*}\). The scaling argument
\(r\mapsto\lambda r\) with \(|\Fp|\geq d^{*}+1\) (which holds since \(d^{*}\leq d\leq p-1<p=|\mathbb{F}_p|\)) separates \(H_c(\lambda r)\) by
degree; the leading component vanishes on \(C(A)^{\perp}\):
\[
  d^{*}\cdot\sum_P \alpha_P^{(d^{*})}c_P\,r_P^{d^{*}-1}=0
  \quad\forall r\in C(A)^{\perp}.
\]

The vanishing identity $d^*\cdot\sum_P\alpha_P^{(d^*)}c_P r_P^{d^*-1}=0$
on $C(A)^\perp$ says that the degree-$(d^*-1)$ homogeneous form
$Q_c(r)=\sum_P\alpha_P^{(d^*)}c_P r_P^{d^*-1}$ vanishes on
$C(A)^\perp$. Since $d^*\leq p-1 < p$, we have $d^*!$ is a unit in
$\mathbb{F}_p$; applying the standard multilinear polarization identity
yields the fully polarized form: for all
$r^{(1)},\dots,r^{(d^*-1)}\in C(A)^\perp$,
\[
  d^{*}!\cdot\sum_P \alpha_P^{(d^{*})}c_P\,
  r_P^{(1)}\,r_P^{(2)}\cdots r_P^{(d^{*}-1)}=0.
\]
The identity says
\(
  \alpha^{(d^{*})}*c\;\in\;
  \bigl((C(A)^{\perp})^{*(d^{*}-1)}\bigr)^{\perp}.
\)
Therefore, by Corollary~\ref{cor:iter-schur},
\((C(A)^{\perp})^{*(d^{*}-1)}=\Fp^G\), so the orthogonal complement is
\(\{0\}\). Thus \(\alpha^{(d^{*})}*c=0\) for every \(c\in C(A)\), and
choosing a character \(c=\chi_{\alpha_0}\) with all coordinates
nonzero forces \(\alpha^{(d^{*})}=0\), contradicting the choice of
\(d^{*}\).

Hence \(d^{*}\leq 2\), as claimed.
\end{proof}

\subsection{Complete classification and the Clifford ceiling}\label{subsec:classification}

\begin{theorem}[Spectral formula for the multiplier algebra]
\label{thm:mult-stab}
Let $G$ be a finite abelian group with $\exp(G)\mid p-1$,
$H=\widehat{G}$, and $A\subseteq H$ nonempty. Define the
\emph{translation stabilizer} and the \emph{multiplier algebra}
$\mathrm{Stab}(A) := \{b\in H : A+b = A\},$ 
$\mathrm{Mult}(C(A)) := \{f\in\mathbb{F}_p^G : f*C(A)\subseteq C(A)\}$.
Then, $$\mathrm{Mult}(C(A)) = C(\mathrm{Stab}(A)).$$
\end{theorem}

\begin{proof}
Write $\beta=\sum_{b\in H}\widehat\beta(b)\chi_b$. For each $\alpha\in A$,
\(\displaystyle 
  \beta*\chi_\alpha
  = \sum_{b\in H}\widehat\beta(b)\,\chi_{\alpha+b}.
\)
This lies in $C(A)$ for all $\alpha\in A$ if and only if
$\widehat\beta(b)\neq 0\Rightarrow A+b\subseteq A$.
Since translation by $b$ is bijective and $A$ is finite,
$A+b\subseteq A\Leftrightarrow A+b=A\Leftrightarrow b\in\mathrm{Stab}(A)$.
Hence $\mathrm{supp}(\widehat\beta)\subseteq\mathrm{Stab}(A)$,
i.e., $\beta\in C(\mathrm{Stab}(A))$.
\end{proof}

Theorems~\ref{thm:cubic} and~\ref{thm:degree-d-nogo} reduce
classification to the quadratic sector. We classify the remaining
sector via the multiplier algebra of the classical code.

\begin{theorem}[Clifford ceiling --- general classification]
\label{thm:clifford-ceiling}
Let $p\geq 5$ be prime, $G$ a finite abelian group with
$\exp(G)\mid p-1$, $H=\widehat{G}$, and $A\subseteq H$
nonempty with $A\cap(-A)=\varnothing$.
For every $2\le d\le p-1$,
\[
  S^{\mathrm{casi}}_{\le d}(A)
  = \mathbb{F}_p^G \;\oplus\; C(\mathrm{Stab}(A)),
\]
and under the strict CSS-preserving condition,
\[
  S^{\mathrm{CSS}}_{\le d}(A)
  = C(A)^\perp \;\oplus\; C(\mathrm{Stab}(A)).
\]
In particular, every strictly CSS-preserving local-diagonal gate
lies at level~$\le 2$ of the Clifford hierarchy.
We call this structural bound the \emph{Clifford ceiling}.
\end{theorem}

\begin{proof}
\emph{Step 1: degrees $\geq 3$ vanish.}
Theorem~\ref{thm:degree-d-nogo} gives $\alpha^{(k)}=0$
for every $k\geq 3$ and every $P\in G$.
 
\emph{Step 2: quadratic sector via degree separation and the multiplier
algebra.}
With $\alpha^{(k)}=0$ for $k\geq 3$, the phase reduces to
$F(x)=\sum_P(\beta_P x_P^2+\gamma_P x_P)$.
Fix $c\in C=C(A)$. Expanding the CASI shift:
\(
  F(r+c)-F(r)
  = \sum_P\bigl(2\beta_P c_P\, r_P + \beta_P c_P^2 + \gamma_P c_P\bigr)
  = 2\langle\beta*c,\,r\rangle + \text{const}(c),
\)
where $\text{const}(c):=\sum_P(\beta_P c_P^2+\gamma_P c_P)$ is
independent of $r$.
By the CASI criterion~\eqref{eq:casi}, the entire expression must be
independent of $r\in C^\perp$; in particular the term linear in $r$
must vanish identically on $C^\perp$.
Substituting $\lambda r$ for $r$ (with $\lambda\in\Fp$) and using that
$C^\perp$ is an $\Fp$-subspace, the linear term scales by $\lambda$
while $\text{const}(c)$ is unchanged; comparing at two distinct $\lambda$
isolates the linear part:
\(
  2\langle\beta*c,\,r\rangle = 0
  \qquad\text{for all }r\in C^\perp.
\)
Since $p\geq 5$ implies $2\neq 0$ in $\Fp$, we conclude
$\langle\beta*c,r\rangle=0$ for all $r\in C^\perp$, i.e.,
$\beta*c\in(C^\perp)^\perp=C$.
Since this holds for every $c\in C$, we have $\beta\in\mathrm{Mult}(C(A))$.
By Theorem~\ref{thm:mult-stab},
\(
  \mathrm{Mult}(C(A))=C(\mathrm{Stab}(A)).
\)
 
\emph{Step 3: linear sector.}
In the CASI setting the linear part $\gamma$ is unconstrained (the shift
$\langle\gamma*c,r\rangle$ is absorbed into $\phi(c)$).
Under strict CSS preservation ($\phi=0$), vanishing of the constant part
$\text{const}(c)=\langle\beta*c+\gamma,c\rangle$ for all $c\in C$, combined
with $\beta*c\in C\subseteq C^\perp$, forces $\gamma\in C^\perp$.
 
\emph{Conclusion.}
The three steps give $S^{\mathrm{casi}}_{\leq d}=\Fp^G\oplus C(\mathrm{Stab}(A))$
(the $\Fp^G$ factor parametrises the unconstrained linear coefficients) and
$S^{\mathrm{CSS}}_{\leq d}=C(A)^\perp\oplus C(\mathrm{Stab}(A))$.
By Remark~\ref{rem:degree-level}, degree~$\leq 2$ equals level~$\leq 2$ of
the Clifford hierarchy.
\end{proof}

\begin{remark}[Degree equals level; completeness within the local-diagonal ansatz]
\label{rem:degree-level}
By Theorem~2 of Cui--Gottesman--Krishna~\cite{CGK17}, for
precision-$1$ diagonal gates over $\mathbb{F}_p$, the Clifford
hierarchy level equals the polynomial degree of the local phase
function. Hence degree~$\leq 2$ is equivalent to level~$\leq 2$,
and the classification of Theorem~\ref{thm:clifford-ceiling} is
complete within the local-diagonal ansatz: for prime~$p$, Lagrange
interpolation shows that every single-site function
$f_P:\mathbb{F}_p\to\mathbb{F}_p$ has a polynomial representative of
degree~$\leq p-1$, so no local-diagonal gate above level~$2$ can exist.
The relation to the Eastin--Knill theorem is discussed in
Section~\ref{subsec:eastin-knill}.
\end{remark}

\subsection{A general Schur obstruction for CSS pairs}
\label{subsec:general-schur-obstruction}

The Schur saturation theorem~\ref{thm:schur} and the
degree-descent argument (Theorem~\ref{thm:degree-d-nogo}) were
established for symmetric spectral codes $Q(A)=\CSS(C(A),C(A))$.
The underlying mechanism, however, depends only on two properties of
the ambient dual space: that its Schur square fills $\mathbb{F}_p^n$,
and that the $X$-check code covers every coordinate.
We isolate these conditions in a single theorem applicable to any CSS
pair.

Let $p\geq 5$ be prime and let $Q=\CSS(C_X,C_Z)$ be a CSS code over
$\mathbb{F}_p^n$ with $C_X\subseteq C_Z^\perp$.
Set $D:=C_Z^\perp$.
A local polynomial phase
$F(x)=\sum_{i=1}^n\sum_{k=1}^{p-1}\alpha_i^{(k)}x_i^k$
satisfies the \emph{CASI criterion with respect to $(C_X,D)$} if, for
every $c\in C_X$, the difference $F(r+c)-F(r)$ is independent of
$r\in D$ (cf.~Definition~\ref{def:casi}).

\begin{theorem}[General Schur obstruction]
\label{thm:general-schur-obstruction}
Let $p\geq 5$ be prime, let $Q=\CSS(C_X,C_Z)$ be a CSS code over
$\mathbb{F}_p^n$ with $C_X\subseteq C_Z^\perp$, and set
$D:=C_Z^\perp$.
Suppose:
\begin{enumerate}[label=\textup{(\roman*)},leftmargin=1.6em]
\item $D*D=\mathbb{F}_p^n$; and
\item $C_X$ has full coordinate support: for every
  $i\in\{1,\dots,n\}$ there exists $c\in C_X$ with $c_i\neq 0$.
\end{enumerate}
Then every local polynomial phase satisfying the CASI criterion
satisfies $\alpha_i^{(k)}=0$ for every $k\geq 3$ and every $i$.
Consequently, every CASI-compatible local diagonal gate lies at
Clifford hierarchy level at most~$2$.
\end{theorem}

\begin{proof}
\emph{Step~1: iteration.}
Condition~(i) forces $D$ to have full coordinate support: if some
$P\in\{1,\dots,n\}$ were zero for all $u\in D$, then $\delta_P\notin
D*D=\mathbb{F}_p^n$, a contradiction.
Full coordinate support gives $\mathbb{F}_p^n*D=\mathbb{F}_p^n$: for
each $P$, pick $u\in D$ with $u_P\neq 0$; then
$u_P^{-1}\delta_P*u=\delta_P$.
Hence $D^{*s}=(D^{*(s-1)})*D\supseteq\mathbb{F}_p^n*D=\mathbb{F}_p^n$
for every $s\geq 2$ by induction.

\emph{Step~2: degree descent.}
Let $d^*=\max\{k:\alpha^{(k)}\neq 0\}$ and suppose $d^*\geq 3$.
For $c\in C_X$, define $H_c(r):=F(r+c)-F(r)-F(c)+F(0)$.
By the CASI criterion, $H_c(r)=0$ for all $r\in D$.
The degree-$(d^*-1)$ component of $H_c(r)$ in $r$ is
$d^*\sum_i\alpha_i^{(d^*)}c_ir_i^{d^*-1}$.
Since $D$ is an $\mathbb{F}_p$-subspace and $d^*\leq p-1$,
substituting $r\mapsto\lambda r$ and separating degrees gives
\[
  \sum_i\alpha_i^{(d^*)}c_ir_i^{d^*-1}=0
  \qquad\forall r\in D.
\]
Since $d^*!\neq 0$ in $\mathbb{F}_p$, multilinear polarization yields
\(
  \alpha^{(d^*)}*c\;\in\;\bigl(D^{*(d^*-1)}\bigr)^\perp=\{0\},
\)
where the last equality uses Step~1.
Hence $\alpha^{(d^*)}*c=0$ for every $c\in C_X$.
By condition~(ii), for each $i$ there exists $c\in C_X$ with
$c_i\neq 0$, so $\alpha_i^{(d^*)}=0$ for all $i$, contradicting
$d^*\geq 3$.
\end{proof}

\begin{remark}[Relation to the spectral case]
\label{rem:general-vs-spectral}
In the symmetric spectral case $C_X=C_Z=C(A)$ with
$A\cap(-A)=\varnothing$, Theorem~\ref{thm:schur} gives
$D=C(A)^\perp$ and $D*D=\mathbb{F}_p^G$ automatically, and
every character $\chi_{\alpha_0}$ with $\alpha_0\in A$ is nowhere
zero on the torus, supplying full coordinate support.
Theorem~\ref{thm:general-schur-obstruction} therefore subsumes
Theorem~\ref{thm:degree-d-nogo}, and its Step~1 provides the
simpler iteration argument for Corollary~\ref{cor:iter-schur}.
\end{remark}

\begin{definition}[Asymmetric multiplier space]
\label{def:asymmetric-multiplier}
For a CSS pair $C_X\subseteq C_Z^\perp$ over $\mathbb{F}_p^n$, define
\[
  \Mult(C_X,C_Z)
  :=
  \bigl\{\beta\in\mathbb{F}_p^n:\beta*C_X\subseteq C_Z\bigr\}.
\]
\end{definition}

When $C_X=C_Z=C(A)$, Theorem~\ref{thm:mult-stab} gives
$\Mult(C(A),C(A))=C(\mathrm{Stab}(A))$, recovering the multiplier
algebra of the symmetric spectral case.

\begin{proposition}[Quadratic sector for CSS pairs]
\label{prop:quadratic-css-pairs}
Under the hypotheses of Theorem~\ref{thm:general-schur-obstruction},
\[
  S^{\mathrm{CASI}}_{\leq 2}(C_X,C_Z)
  =\mathbb{F}_p^n\oplus\Mult(C_X,C_Z),
  \qquad
  S^{\mathrm{CSS}}_{\leq 2}(C_X,C_Z)
  =C_X^\perp\oplus\Mult(C_X,C_Z),
\]
where the first summand parametrises linear coefficients and the second
quadratic coefficients.
\end{proposition}

\begin{proof}
All degrees $\geq 3$ vanish by Theorem~\ref{thm:general-schur-obstruction},
so $F(x)=\sum_i\beta_ix_i^2+\sum_i\gamma_ix_i$.
For $c\in C_X$ and $r\in D=C_Z^\perp$,
\(
  F(r+c)-F(r)
  =2\langle\beta*c,\,r\rangle+\sum_i\beta_ic_i^2+\langle\gamma,c\rangle.
\)
Independence of $r$ requires $\langle\beta*c,r\rangle=0$ for all
$r\in C_Z^\perp$, i.e., $\beta*c\in C_Z$.
Since this holds for all $c\in C_X$, we have $\beta\in\Mult(C_X,C_Z)$.
The linear part $\gamma$ is unconstrained under CASI.
Under strict CSS preservation, the constant part must also vanish:
since $\beta*c\in C_Z$ and $c\in C_X\subseteq C_Z^\perp$, the CSS
condition gives $\langle\beta*c,c\rangle=\sum_i\beta_ic_i^2=0$,
so $\langle\gamma,c\rangle=0$ for all $c\in C_X$, hence
$\gamma\in C_X^\perp$.
\end{proof}

\begin{definition}[Schur defect spaces]
\label{def:schur-defect}
For a CSS pair $C_X\subseteq C_Z^\perp$ and $d\geq 3$, the
\emph{degree-$d$ Schur defect space} is
\[
  \mathcal{A}_d(C_X,C_Z)
  :=
  \bigl\{\alpha\in\mathbb{F}_p^n:
  \alpha*C_X
  \subseteq\bigl((C_Z^\perp)^{*(d-1)}\bigr)^\perp\bigr\}.
\]
\end{definition}

\begin{proposition}[Necessary Schur condition]
\label{prop:necessary-schur}
If $F$ satisfies the CASI criterion and has leading degree
$d^*\geq 3$, then $\alpha^{(d^*)}\in\mathcal{A}_{d^*}(C_X,C_Z)$.
In particular, $\mathcal{A}_d(C_X,C_Z)=0$ implies no CASI-compatible
phase can have leading degree $d$.
If additionally $(C_Z^\perp)^{*(d-1)}=\mathbb{F}_p^n$ and $C_X$
has full coordinate support, then $\mathcal{A}_d(C_X,C_Z)=0$.
\end{proposition}

\begin{proof}
The degree-descent step of Theorem~\ref{thm:general-schur-obstruction}
gives $\alpha^{(d^*)}*c\in\bigl(D^{*(d^*-1)}\bigr)^\perp$ for every
$c\in C_X$, i.e., $\alpha^{(d^*)}\in\mathcal{A}_{d^*}(C_X,C_Z)$.
If $(C_Z^\perp)^{*(d-1)}=\mathbb{F}_p^n$, then
$\bigl((C_Z^\perp)^{*(d-1)}\bigr)^\perp=\{0\}$, and full coordinate
support of $C_X$ forces $\alpha^{(d)}=0$.
\end{proof}

\paragraph{Non-symmetric spectral pairs.}
Consider a spectral CSS pair\linebreak $Q=\CSS(C(A_X),C(A_Z))$ such that
$A_X\cap(-A_Z)=\varnothing$, the dual space is
$D=C(A_Z)^\perp=C(B_Z)$ where $B_Z:=H\setminus(-A_Z)$, and
$D*D=C(B_Z+B_Z)$.
Schur saturation holds if and only if $B_Z+B_Z=H$.
In the symmetric case $A_X=A_Z=A$, the condition $A\cap(-A)=\varnothing$
forces $|B|>|H|/2$, and the majority-sum lemma gives $B+B=H$
automatically.
In the non-symmetric case, $A_X\cap(-A_Z)=\varnothing$ imposes no
lower bound on $|B_Z|$; if $|B_Z|\leq|H|/2$, then $B_Z+B_Z$ may be
a proper subgroup of $H$, Schur saturation fails, and
$\mathcal{A}_d(C_X,C_Z)$ may be nonzero.
This is the mechanism behind the escape from the ceiling in
non-symmetric constructions (Section~\ref{subsec:escape}).

\paragraph{LDPC specialization.}
For CSS LDPC codes, condition~(ii) of
Theorem~\ref{thm:general-schur-obstruction} holds whenever every
column of $H_X$ is nonzero, the standard non-degeneracy requirement.
Condition~(i), $(\ker H_Z)*(\ker H_Z)=\mathbb{F}_p^n$, must be
verified case by case; Lemma~\ref{lem:schur-criterion} in
Section~\ref{subsec:ldpc-hgp} gives an exact algebraic criterion
and a concrete verified instance.

\section{Applications and Scope of the Clifford Ceiling}
\label{sec:applications}

\subsection{Symmetric Reed-Solomon CSS codes}\label{subsec:rs}

Let $p\geq 5$ be prime and set $G=\mathbb{F}_p^*$, the multiplicative
group of $\mathbb{F}_p$, cyclic of order $n=p-1$. Its character group
satisfies $\widehat{G}\cong\mathbb{Z}_n$ via $\chi_j(g)=g^j$, so
$\exp(G)=n=p-1$ divides $p-1$. For $1\leq k<n/2$, let
$A=\{1,2,\ldots,k\}\subseteq\mathbb{Z}_n$. The spectral code
$C(A)=\mathrm{span}_{\mathbb{F}_p}\{g\mapsto g^j:1\leq j\leq k\}$
is the Reed-Solomon code of length $n$ and dimension $k$ obtained by
evaluating the monomials $\{x,x^2,\ldots,x^k\}$ at all nonzero
elements of $\mathbb{F}_p$; it is MDS with minimum distance $p-k$.

\begin{proposition}[Symmetric RS Clifford ceiling]
\label{prop:rs-ceiling}
Let $p\geq 5$ be prime and $1\leq k<(p-1)/2$. The symmetric
Reed-Solomon CSS code $Q(A)=\mathrm{CSS}(C(A),C(A))$ with
$A=\{1,\ldots,k\}$ has parameters $[[p-1,\,p-1-2k,\,k+1]]_p$
and satisfies the Clifford ceiling:
\[
  S^{\mathrm{CSS}}_{\leq p-1}(A) = C(A)^\perp \oplus \langle\Qunif\rangle.
\]
In particular, no strictly CSS-preserving local-diagonal transversal
gate lies above level~$2$ of the Clifford hierarchy.
\end{proposition}

\begin{proof}
\emph{Self-orthogonality.} In $\mathbb{Z}_n$, negation maps
$j\mapsto n-j$, so $-A=\{n-k,\ldots,n-1\}$. Since $k<n/2$,
the minimum of $-A$ is $n-k>k$, giving $A\cap(-A)=\varnothing$;
by Corollary~\ref{cor:selforth}, $C(A)\subseteq C(A)^\perp$
and $Q(A)$ is well-defined.

\emph{Schur saturation.} Since $\exp(G)=p-1$ divides $p-1$,
Theorem~\ref{thm:schur} gives $C(A)^\perp*C(A)^\perp=\mathbb{F}_p^G$.

\emph{Stabilizer.} The set $A=\{1,\ldots,k\}$ is a contiguous
interval in $\mathbb{Z}_n$ not wrapping around (since $k<n/2<n$).
Any $b\in\mathrm{Stab}(A)$ must satisfy
$\{1+b,\ldots,k+b\}=\{1,\ldots,k\}$ modulo $n$, which forces
$b=0$. Hence $\mathrm{Stab}(A)=\{0\}$, and
Theorem~\ref{thm:clifford-ceiling} gives
$S^{\mathrm{CSS}}=C(A)^\perp\oplus C(\{0\})
=C(A)^\perp\oplus\langle\Qunif\rangle$.

\emph{Parameters.} The dual $C(A)^\perp=C(B)$ with
$B=\{0,1,\ldots,p-k-2\}$ is the MDS code $\mathrm{RS}(p-1,p-k-1)$
with minimum distance $k+1$. Since every nonzero codeword of $C(A)$
has weight $\geq p-k>k+1$, the minimum-weight elements of
$C(A)^\perp\setminus C(A)$ have weight $k+1$, giving $d_Q=k+1$.
\end{proof}

\begin{remark}
To our knowledge, Proposition~\ref{prop:rs-ceiling} is the first
explicit gate-group classification for symmetric Reed-Solomon CSS
codes. The non-symmetric constructions of~\cite{HE25} achieve
transversal non-Clifford gates precisely by breaking the symmetry
condition; see Section~\ref{subsec:escape}.
\end{remark}

\begin{example}[Symmetric RS code, false-positive at level~5]
\label{ex:rs-p13}
Let $p=13$, $k=2$, and $n=p-1=12$.
The support $A=\{1,2\}\subseteq\mathbb{Z}_{12}$ satisfies
$A\cap(-A)=\{1,2\}\cap\{10,11\}=\varnothing$, so the symmetric CSS code
$Q(A)=\mathrm{CSS}(\mathrm{RS}(12,2),\mathrm{RS}(12,2))$ is well-defined with
parameters $[[12,8,3]]_{13}$.
 
The $\ell$-orthogonality threshold is $\ellmax=5$.
Indeed, for $\ell\leq 5$ the Minkowski sum $(\ell-1)A=\{\ell-1,\ldots,2(\ell-1)\}$
satisfies $\max((\ell-1)A)=2(\ell-1)\leq 8<10=\min(-A)$, so
$((\ell-1)A)\cap(-A)=\varnothing$ and $C(A)$ is $\ell$-orthogonal.
At $\ell=6$ the element $5\cdot 2=10$ lies in both $(5A)$ and $-A=\{10,11\}$,
so $\ell$-orthogonality fails.
 
By the spectral criterion (Theorem~\ref{thm:criterion}), $\ellmax=5$
certifies that $Q(A)$ passes the orthogonality-type screening for a transversal
diagonal gate candidate at level~$5$ of the Clifford hierarchy: for $p=13$
this points to a degree-$5$ non-Clifford phase.
Since $\mathrm{Stab}(A)=\{0\}$ (the interval $\{1,2\}$ is aperiodic in
$\mathbb{Z}_{12}$), Proposition~\ref{prop:rs-ceiling} gives
$S^{\mathrm{CSS}}_{\leq 12}(A)=C(A)^\perp\oplus\langle\Qunif\rangle$:
the entire non-Clifford sector vanishes, and the surviving gate group collapses
to level~$\leq 2$, regardless of the $\ellmax=5$ certificate.
\end{example}

\subsection{BCH and cyclic codes}\label{subsec:bch}

The same criterion applies to any cyclic code whose defining set
satisfies the anti-symmetry condition. Let $n\mid p-1$, and let
$G$ be the cyclic group of order $n$ with $\widehat{G}\cong\mathbb{Z}_n$;
since $n\mid p-1$ we have $\exp(G)\mid p-1$. A cyclic code of
length $n$ over $\mathbb{F}_p$ with defining set $A\subseteq\mathbb{Z}_n$
is precisely the spectral code $C(A)$ in this framework; see
Camps-Moreno et al.~\cite{CLMRS24} for the CSS-T perspective
on cyclic codes.

\begin{proposition}[Symmetric cyclic and BCH Clifford ceiling]
\label{prop:bch-ceiling}
Let $p\geq 5$ be prime, $n\mid p-1$, and $A\subseteq\mathbb{Z}_n$
nonempty with $A\cap(-A)=\varnothing$. The symmetric cyclic CSS
code $Q(A)=\mathrm{CSS}(C(A),C(A))$ satisfies
\[
  S^{\mathrm{CSS}}_{\leq p-1}(A)=C(A)^\perp\oplus C(\mathrm{Stab}(A)).
\]
In particular, for any narrow-sense BCH code with defining set
$A=\{1,\ldots,\delta-1\}$ and $\delta-1<n/2$, we have
$\mathrm{Stab}(A)=\{0\}$ and
\[
  S^{\mathrm{CSS}}_{\leq p-1}(A)=C(A)^\perp\oplus\langle\Qunif\rangle.
\]
\end{proposition}

\begin{proof}
Since $\exp(G)=n\mid p-1$, Theorem~\ref{thm:schur} gives
$C(A)^\perp*C(A)^\perp=\mathbb{F}_p^G$, and
Theorem~\ref{thm:clifford-ceiling} yields the first formula.
For BCH codes: $A=\{1,\ldots,\delta-1\}$ is a contiguous interval
not wrapping in $\mathbb{Z}_n$ (as $\delta-1<n/2$), so
$\mathrm{Stab}(A)=\{0\}$ by the argument of
Proposition~\ref{prop:rs-ceiling}, giving the second formula.
\end{proof}

\subsection{Boolean toric codes: the closed-form instance}\label{subsec:boolean-toric}

\begin{lemma}[Boolean structure]
\label{lem:structure}
For $p\geq 5$, $A\subseteq\{0,1\}^m\setminus\{0\}$ and any
$\ell\geq 1$:
\begin{enumerate}[label=\textup{(\roman*)},leftmargin=1.6em]
\item $\ell A\subseteq\{0,1,\dots,\ell\}^m$ in $\mathbb{Z}_{p-1}^m$;
\item every $w\in -A$ has at least one coordinate equal to $p-2\geq 3$;
\item $A\cap(-A)=\emptyset$, so by Theorem~\ref{thm:criterion}
  $C(A)\subseteq C(A)^{\perp}$ and $\mathcal{Q}(A)$ is well-defined.
\end{enumerate}
\end{lemma}

\begin{proof}
(i) is immediate from the definition of sumsets. For (ii): in $\mathbb{Z}_{p-1}$, negation maps $0\mapsto 0$ and $1\mapsto (p-1)-1 = p-2$. Since $\alpha\in A$ excludes the zero vector, it has at least one coordinate equal to~$1$. Consequently, $-\alpha$ has at least one coordinate equal to $p-2$. Since $p \geq 5$, we have $p-2 \geq 3$. (iii) follows because the coordinates of elements in $A$ are strictly bounded by $1$, while every element in $-A$ contains a coordinate of at least $3$. The sets are disjoint, so $A\cap(-A)=\emptyset$; applying Corollary~\ref{cor:selforth} (or equivalently, Theorem~\ref{thm:criterion} with $\ell=2$) guarantees $C(A)\subseteq C(A)^\perp$.
\end{proof}

\begin{definition}[Full Boolean toric CSS code]
\label{def:full-boolean-toric}
Let $p\geq 5$ be prime and $m\geq 2$. For any nonempty
$A\subseteq\{0,1\}^m\setminus\{0\}$, the symmetric CSS code
$Q(A)=\mathrm{CSS}(C(A),C(A))$ is called a
\emph{Full Boolean toric CSS code}. By Lemma~\ref{lem:structure}(iii),
$Q(A)$ is well-defined for every such $A$.
\end{definition}

\begin{theorem}[Sharp threshold]
\label{thm:sharp}
Let $p \ge 5$ be a prime, $m \ge 2$, and
$\emptyset \ne A \subseteq \{0,1\}^m \setminus \{0\}$.
For $2 \le \ell \le p-1$:
$$
  C(A) \text{ is } \ell\text{-orthogonal} \iff \ell \le p-2.
$$
The maximal level $\ell_{\max} = p-2$ is independent of $A$.
\end{theorem}

\begin{proof}
We evaluate the intersection $((\ell-1)A) \cap (-A)$ across the
domain of $\ell$.

\textbf{Case $\ell \le p-2$:}
By Lemma~\ref{lem:structure}(i), every vector $v \in (\ell-1)A$
has coordinate values bounded by $\ell-1$. Since $\ell \le p-2$,
the maximum possible value of any coordinate in $v$ is $p-3$.
Simultaneously, by Lemma~\ref{lem:structure}(ii), every vector
$w \in -A$ is guaranteed to have at least one coordinate exactly
equal to $p-2$.
Since $p-3 < p-2$, the vector $v$ can never match $w$ on that
specific coordinate. Therefore, $v \neq w$ universally, meaning
$((\ell-1)A) \cap (-A) = \emptyset$. By
Theorem~\ref{thm:criterion}, $C(A)$ is $\ell$-orthogonal.

\textbf{Case $\ell = p-1$:}
We must show that at $\ell = p-1$, the intersection is no longer
empty. Pick any arbitrary element $\alpha \in A$. Consider the
vector formed by adding $\alpha$ to itself $p-2$ times:
$(p-2)\alpha \in (p-2)A = ((\ell-1)A)$.
We evaluate this scalar multiplication coordinate-by-coordinate
modulo $p-1$. Since $\alpha$ is a Boolean vector, its coordinates
are exclusively $0$ or $1$. If $\alpha_i = 0$: $(p-2)\cdot 0 = 0 \equiv 0 \pmod{p-1}$. If $\alpha_i = 1$: $(p-2)\cdot 1 = p-2 \equiv -1 \pmod{p-1}$.

Notice that multiplying the Boolean vector $\alpha$ by $(p-2)$
yields the exact same coordinate-wise result as negating $\alpha$
modulo $p-1$. Thus, we have the identity:
$
  (p-2)\alpha \equiv -\alpha \pmod{p-1}.
$ 
Since $(p-2)\alpha \in ((\ell-1)A)$ and $-\alpha \in -A$, the
element $-\alpha$ explicitly exists in both sets. The intersection
$((\ell-1)A) \cap (-A)$ is non-empty, and by
Theorem~\ref{thm:criterion}, $\ell$-orthogonality strictly fails
at $\ell = p-1$.
\end{proof}

\begin{definition}[Difference group]
\(\DA:=\langle\alpha-\beta:\alpha,\beta\in A\rangle\leq\Zm\).
\end{definition}

\begin{lemma}[Primal weight bound]
\label{lem:primal-weight}
For every \(p\geq 5\), \(m\geq 2\), and
\(A\subseteq\{0,1\}^m\setminus\{0\}\), every nonzero \(g\in C(A)\)
satisfies \(\wt(g)>3\).
\end{lemma}

\begin{proof}
\emph{Case \(m=2\).} Then \(A\subseteq\{(1,0),(0,1),(1,1)\}\), so any
\(g\in C(A)\) is a Laurent polynomial \(g(x_1,x_2)=a x_1+b x_2+c x_1
x_2\) on \((\Fps)^2\). We bound the number of zeros of \(g\) on
\((\Fps)^2\) by case analysis on the coefficients:
\begin{itemize}[leftmargin=1.5em]
\item If exactly one of \(a,b,c\) is nonzero, \(g\) is a nowhere-zero
character, so \(\wt(g)=(p-1)^2\geq 9\).
\item If \(c=0\) and \(a,b\neq 0\), the equation \(a x_1+b x_2=0\)
gives \(x_2=-(a/b)x_1\), one zero per nonzero \(x_1\); hence
\(\wt(g)=(p-1)^2-(p-1)=(p-1)(p-2)\geq 6\).
\item If \(c\neq 0\) and \(a\neq 0\): we split on \(b\). If \(b=0\), then \(g=x_1(a+cx_2)\), which vanishes precisely on \(\{x_2=-a/c\}\times\mathbb{F}_p^*\), giving exactly \(p-1\) zeros and \(\wt(g)=(p-1)^2-(p-1)=(p-1)(p-2)\geq 6\). If \(b\neq 0\), factor as \(g=x_1(a+cx_2)+bx_2\); the zero set \(\{x_2=-ax_1/(b+cx_1)\}\) requires \(b+cx_1\neq 0\) and \(x_1\neq 0\), and yields \(x_2\neq 0\), giving at most \(p-2\) zeros and \(\wt(g)\geq(p-1)^2-(p-2)\geq 7\). In both subcases \(\wt(g)\geq 6>3\).
\item If \(a=0\) and \(b,c\neq 0\): \(g=x_2(b+c x_1)\) vanishes on
\(\{x_1=-b/c\}\times\Fps\), so \(\wt(g)=(p-1)^2-(p-1)=(p-1)(p-2)\geq 6\).
\end{itemize}
In every case \(\wt(g)\geq(p-1)(p-2)\geq 6>3\) for \(p\geq 5\).

\emph{Case \(m\geq 3\).} By the
Donoho--Stark--Meshulam uncertainty principle for finite abelian
groups~\cite{Mesh06},
\[
  \wt(g)\cdot|\supp(\widehat g)|\geq |G|=(p-1)^m.
\]
Since \(\widehat g\) is supported on \(A\subseteq\{0,1\}^m\setminus\{0\}\),
\(|\supp(\widehat g)|\leq |A|\leq 2^m-1\), so
\[
  \wt(g)\geq \frac{(p-1)^m}{2^m-1}\geq \frac{3^m}{2^m-1}.
\]
For \(m=3\), \(3^3/(2^3-1)=27/7>3\). The function
\(m\mapsto 3^m/(2^m-1)\) is strictly increasing for \(m\geq 1\)
(derivative analysis or direct ratio: \(3^{m+1}/(2^{m+1}-1)\) divided by
\(3^m/(2^m-1)\) is \(3(2^m-1)/(2^{m+1}-1)>1\) iff \(2^{m+1}>2\), which
always holds), so \(\wt(g)>3\) for all \(m\geq 3\).
\end{proof}

\begin{theorem}[Distance dichotomy and determinantal criterion]
\label{thm:distance}
Let $p\geq 5$, $m\geq 2$, and $A\subseteq\{0,1\}^m\setminus\{0\}$.
Then $\QA$ is well-defined and the following hold.
 
\emph{(a) Dichotomy.}
\[
  \dQ(\QA)\in\{2,3\},\qquad \dQ(\QA)=3\iff\DA=\Zm.
\]
 
\emph{(b) Determinantal criterion for minimal supports.}
When $|A|=m+1$, write $A=\{\alpha_0,\alpha_1,\dots,\alpha_m\}$ and
\[
  M_A := \bigl[\alpha_1-\alpha_0\mid\cdots\mid\alpha_m-\alpha_0\bigr]
       \in M_m(\mathbb{Z}).
\]
The following conditions are mutually equivalent:
\textup{(i)} $\dQ(\QA)=3$;\quad
\textup{(ii)} $\DA=\Zm$;\quad
\textup{(iii)} $\det(M_A)$ is a unit in $\mathbb{Z}_{p-1}$;\quad
\textup{(iv)} $\gcd(\det M_A,\,p-1)=1$.
\end{theorem}
 
\begin{proof}
Well-definedness is Lemma~\ref{lem:structure}(iii).
 
\medskip\noindent\textbf{Proof of part~(a).}
 
\emph{Step 1 (projectivity): $\dmin(\CAp)\geq 3\Longleftrightarrow\DA=\Zm$.}
 
\emph{($\Rightarrow$, contrapositive.)}
Suppose $\DA\neq\Zm$; pick $\lambda\in\Zm\setminus\{0\}$ with
$\lambda\cdot(\alpha-\beta)=0$ for all $\alpha,\beta\in A$.
Then $\lambda\cdot\alpha$ is constant, say equal to $D\in\mathbb{Z}_{p-1}$,
on $A$.
Let $\xi\in\Fps$ be a generator, fix any $a\in G$, and set
$b_i=\xi^{\lambda_i}a_i$.
Since some $\lambda_i\not\equiv 0$, we have $b\neq a$.
The element $f=\delta_a-\xi^{-D}\delta_b\in\Fp^G$ has weight~$2$, and for
every $\alpha\in A$,
\(
  \langle f,\chi_\alpha\rangle
  =a^\alpha-\xi^{-D}\xi^{\lambda\cdot\alpha}a^\alpha
  =a^\alpha(1-\xi^0)=0,
\)
so $f\in\CAp$ and $\dmin(\CAp)\leq 2$.
 
\emph{($\Leftarrow$.)}
Suppose $\DA=\Zm$ and $f=c_1\delta_a+c_2\delta_b\in\CAp$ with
$a\neq b$, $c_1,c_2\in\Fps$.
Orthogonality to every $\chi_\alpha$ gives $(b/a)^\alpha=-c_1/c_2$ for
all $\alpha\in A$.
Subtracting for $\alpha,\beta\in A$ yields $(b/a)^{\alpha-\beta}=1$ for
every generator of $\DA=\Zm$, forcing $b=a$, a contradiction.
Hence $\dmin(\CAp)\geq 3$.
 
\emph{Step 2 (universal upper bound): $\dmin(\CAp)\leq 3$ for every $A$.}
Choose three distinct $t_1,t_2,t_3\in\Fps$ and set
$P_j=(t_j,1,\ldots,1)$, $c_1=t_2-t_3$, $c_2=t_3-t_1$, $c_3=t_1-t_2$.
The weight-$3$ element $f=\sum_j c_j\delta_{P_j}$ satisfies
$\langle f,\chi_\alpha\rangle=\sum_j c_j t_j^{\alpha_1}=0$ for
$\alpha_1\in\{0,1\}$, so $f\in\CAp$.
 
\emph{Step 3 (CSS distance).}
Combining Steps~1--2 with Lemma~\ref{lem:primal-weight}:
if $\DA\neq\Zm$, a weight-$2$ element $f\in\CAp\setminus C(A)$ gives
$\dQ=2$; if $\DA=\Zm$, the weight-$3$ bound gives $\dQ=3$.
 
\medskip\noindent\textbf{Proof of part~(b).}
The equivalence (i)$\Leftrightarrow$(ii) is part~(a).
The equivalence (iii)$\Leftrightarrow$(iv) is the definition of unit.
For (ii)$\Leftrightarrow$(iii): $\DA$ is generated by the columns of
$M_A$; surjectivity of $M_A:\mathbb{Z}_{p-1}^m\to\mathbb{Z}_{p-1}^m$
(equivalently bijectivity, since the domain is finite) holds if and only if
$\gcd(\det M_A,p-1)=1$.
\end{proof}

\begin{definition}[Rate-optimal support]
For \(m\geq 3\) and \(p\geq 5\),
\[
  A_m^{*}:=\{e_1,\dots,e_m,\,e_1+e_2\}\subseteq\{0,1\}^m,
\]
and \(Q_m^{*}:=Q(A_m^{*})\).
\end{definition}

\begin{theorem}[Rate-optimal family]
\label{thm:optimal}
For \(p\geq 5\) and \(m\geq 3\):
\begin{enumerate}[label=\textup{(\roman*)},leftmargin=1.6em]
\item \(Q_m^{*}=[[(p-1)^m,\,(p-1)^m-2(m+1),\,3]]_p\);
\item \(\ellmax(A_m^{*})=p-2\);
\item among all \(A\subseteq\{0,1\}^m\setminus\{0\}\) with
  \(\dQ(\QA)=3\), \(Q_m^{*}\) maximizes the logical dimension.
\end{enumerate}
\end{theorem}

\begin{proof}
With \(\alpha_0=e_1+e_2\) and \(\alpha_i=e_i\), the matrix of
differences is \(M_{A_m^{*}}=I_m-(e_1+e_2)\mathbf{1}^\top\). The
matrix determinant lemma gives \(\det M_{A_m^{*}}=1-2=-1\), and
\(\gcd(-1,p-1)=1\); thus \(\dQ=3\) by
Theorem~\ref{thm:distance}(b). Part~(ii) is Theorem~\ref{thm:sharp}.
For (iii), \(\dQ=3\) requires \(\DA=\Zm\)
(Theorem~\ref{thm:distance}(a)), a group of rank~\(m\); since \(\DA\) is
generated by \(|A|-1\) differences \(\alpha-\alpha_0\), we need
\(|A|-1\geq m\), i.e., \(|A|\geq m+1\). Since
\(k=(p-1)^m-2|A|\) is maximized by \(|A|=m+1\) and
\(|A_m^{*}|=m+1\), the family is rate-optimal.
\end{proof}

\begin{example}[Flagship: \(p=5\), \(m=3\)]
\label{ex:flagship}
For \(p=5\), \(m=3\), the rate-optimal support
\(A_3^{*}=\{e_1,e_2,e_3,e_1+e_2\}\) gives
\(Q_3^{*}=[[64,56,3]]_5\). The matrix of differences with
\(\alpha_0=e_1+e_2\) is \(M_{A_3^{*}}=I_3-(e_1+e_2)\mathbf{1}^\top\),
so \(\det M_{A_3^{*}}=-1\), \(\gcd(-1,4)=1\), and
\(\dQ=3\). With $\ellmax=3$, the complete local-diagonal classification gives:
\(\dim S^{\mathrm{casi}}=65\), \(\dim S^{\mathrm{CSS}}=61\),
derived from the theorems in Remark~\ref{rem:numerics}.
\end{example}

\begin{corollary}[Boolean toric Clifford ceiling]
\label{cor:boolean-ceiling}
Let $p\geq 5$ be prime, $m\geq 2$, and
$A\subseteq\{0,1\}^m\setminus\{0\}$ with $\dQ(\QA)=3$.
Then $\mathrm{Stab}(A)=\{0\}$, so $C(\mathrm{Stab}(A))=\langle\Qunif\rangle$,
and the Clifford ceiling of Theorem~\ref{thm:clifford-ceiling} gives
\[
  S^{\mathrm{casi}}_{\le d}(A)
  = \mathbb{F}_p^G\;\oplus\;\langle\Qunif\rangle,
  \qquad
  S^{\mathrm{CSS}}_{\le d}(A)
  = C(A)^\perp\;\oplus\;\langle\Qunif\rangle,
\]
for every $2\le d\le p-1$, where $\Qunif(x)=\sum_{P\in G}x_P^2$.
No gate in $S^{\mathrm{CSS}}$ lies above level~$2$ of the Clifford
hierarchy.
\end{corollary}

\begin{proof}
Since $d_Q(\QA)=3$, Theorem~\ref{thm:distance}(a) gives $\DA=\Zm$. Suppose $b\in\mathrm{Stab}(A)$ with $b\neq 0$, and let $i$ be a coordinate with $b_i\neq 0$. Since $A\subseteq\{0,1\}^m$ and $\DA=\Zm$, both values $0$ and $1$ appear in the projection of $A$ onto coordinate $i$, so there exist $a,a'\in A$ with $a_i=0$ and $a'_i=1$. Then $(a+b)_i=b_i$ and $(a'+b)_i=1+b_i$ must both lie in $\{0,1\}$ in $\Z_{p-1}$. This forces $b_i=0$, contradicting $b_i\neq 0$. Hence $\mathrm{Stab}(A)=\{0\}$, and the result follows from Theorem~\ref{thm:clifford-ceiling} with
$C(\mathrm{Stab}(A))=C(\{0\})=\langle\Qunif\rangle$.
\end{proof}

The only non-Pauli gate surviving the Clifford ceiling is
$\Qunif$. Its survival does not contradict the rigidity claim:
$\Qunif$ is a diagonal Clifford gate and carries no magic.
The following proposition makes its logical action precise.

\begin{proposition}[Logical action of $\Qunif$]
\label{prop:qunif-logical}
Under the hypotheses of Corollary~\ref{cor:boolean-ceiling},
the gate $\Qunif$ descends to a well-defined nontrivial diagonal
Clifford operation on the logical space $C(A)^\perp/C(A)$,
whose associated polar form is nondegenerate.
In particular, for $Q_3^* = [[64,56,3]]_5$, the induced logical
quadratic form has rank~$56$.
\end{proposition}

\begin{proof}
Write $C := C(A)$, $C^{\perp} := C(A)^{\perp}$, and define the map
\[
  q_{\mathrm{unif}}\colon C^{\perp}/C \longrightarrow \mathbb{F}_p,
  \qquad
  q_{\mathrm{unif}}([r]) := \sum_{P \in G} r_P^2.
\]

\emph{Step 1: Well-definedness.}
Let $r \in C^{\perp}$ and $c \in C$. Expanding the square:
\[
  q_{\mathrm{unif}}(r + c) - q_{\mathrm{unif}}(r)
  = 2\sum_{P \in G} r_P c_P + \sum_{P \in G} c_P^2
  = 2\langle r, c \rangle + \langle c, c \rangle.
\]
The first term vanishes because $r \in C^{\perp}$ and $c \in C$.
The second vanishes because $C \subseteq C^{\perp}$.
Hence $q_{\mathrm{unif}}([r])$ is independent of the choice of
representative, and $q_{\mathrm{unif}}$ is a well-defined
quadratic form on $C^{\perp}/C$.

\emph{Step 2: The polar form.}
Polarising $q_{\mathrm{unif}}$ yields the associated symmetric bilinear form:
\[
  B_{\mathrm{unif}}([r],[s])
  := q_{\mathrm{unif}}([r + s]) - q_{\mathrm{unif}}([r]) 
     - q_{\mathrm{unif}}([s])
  = 2\sum_{P \in G} r_P s_P
  = 2\langle r, s \rangle.
\]

\emph{Step 3: Nondegeneracy.}
Suppose $[r]$ lies in the radical of $B_{\mathrm{unif}}$, that is,
\(
  B_{\mathrm{unif}}([r],[s]) = 2\langle r, s \rangle = 0
  \ \text{for every } s \in C^{\perp}.
\)
Since $p \geq 5$, we have $2 \neq 0$ in $\mathbb{F}_p$, so
$\langle r, s \rangle = 0$ for all $s \in C^{\perp}$, i.e.,
$r \in (C^{\perp})^{\perp} = C$.
Hence $[r] = 0$ in $C^{\perp}/C$, proving $B_{\mathrm{unif}}$ is
nondegenerate.

\emph{Step 4: Logical nontriviality and Clifford hierarchy level.}
The logical dimension of $Q(A)$ is positive by assumption, so 
$C^{\perp}/C$ is nonzero. Since $B_{\mathrm{unif}}$ is 
nondegenerate on a nonzero space, the quadratic form 
$q_{\mathrm{unif}}$ is nontrivial, and $\Qunif$ induces a 
nontrivial logical operation. By Theorem~2 of 
Cui--Gottesman--Krishna~\cite{CGK17}, the phase 
$\Qunif(x) = \sum_{P \in G} x_P^2$ has polynomial degree~$2$ in each 
local coordinate and therefore sits at level~$2$ of the Clifford 
hierarchy: it is a diagonal Clifford gate, carrying no magic.

\emph{Step 5: Rank for $Q_3^{*} = [[64,56,3]]_5$.}
For $p = 5$, $m = 3$, $A = A_3^{*}$ with $|A| = m + 1 = 4$:
\[
  |G| = (p-1)^m = 64,
  \,
  \dim C(A) = 4,
  \,
  \dim C(A)^{\perp} = 60,
  \,
  \dim\!\bigl(C(A)^{\perp}/C(A)\bigr) = 56.
\]
Since $B_{\mathrm{unif}}$ is nondegenerate on $C(A)^{\perp}/C(A)$,
the induced logical quadratic form has rank~$56$.
\end{proof}

\begin{corollary}[Prime-$p$ complete local rigidity / No-go for CAB-type distillation blocks]
\label{cor:prime-p}
For every prime $p\geq 5$ and every nonempty
$A\subseteq\{0,1\}^m\setminus\{0\}$ with $d_Q(Q(A))=3$, the complete
CASI local diagonal sector is
\[
  S^{\mathrm{casi}}_{\leq p-1}(A)=\mathbb{F}_p^G\oplus\langle\Qunif\rangle,
\]
while the complete strictly CSS-preserving sector is
\[
  S^{\mathrm{CSS}}_{\leq p-1}(A)=C(A)^\perp\oplus\langle\Qunif\rangle.
\]
In particular, no strictly CSS-preserving local diagonal gate sits above
level~$2$ of the Clifford hierarchy, regardless of how large
$\ell_{\max}=p-2$ is.
\end{corollary}

\begin{proof}
For prime $p$, every function $f:\mathbb{F}_p\to\mathbb{F}_p$
is represented by a unique polynomial of degree $\leq p-1$
(Lagrange interpolation), so the classification of
Corollary~\ref{cor:boolean-ceiling} with $d=p-1$ covers
all local diagonal gates exhaustively.
\end{proof}

\begin{remark}[Explicit verification for $Q_3^*$]
\label{rem:numerics}
For $p=5$, $m=3$, $A=A_3^*$: $|G|=64$, $\dim C(A)=4$,
$\dim C(A)^\perp=60$. Degrees~$\geq 3$ vanish by
Theorems~\ref{thm:cubic} and~\ref{thm:degree-d-nogo};
the degree-$2$ sector contributes exactly $\langle\Qunif\rangle$
by Corollary~\ref{cor:boolean-ceiling}. Hence
\[
  \dim S^{\mathrm{casi}} = 64+1 = 65,
  \qquad
  \dim S^{\mathrm{CSS}} = 60+1 = 61.
\]
For prime~$p$, Lagrange interpolation makes this exhaustive over
all local diagonal gates in the computational basis.
\end{remark}

\subsection{The boundary of the ceiling: non-symmetric codes}
\label{subsec:escape}

The Clifford ceiling rests on two structural properties that break in
the non-symmetric setting, and a concrete example shows exactly where
the argument fails.
 
\paragraph{The two structural triggers.}
First, the symmetry condition $C_1=C_2=C(A)$ forces
$A\cap(-A)=\varnothing$, which gives $|B|=|H\setminus(-A)|>|H|/2$ and
triggers the majority-sum lemma (Lemma~\ref{lem:majority}).
For a non-symmetric code $\mathrm{CSS}(C(A_1),C(A_2))$ with
$C_1\neq C_2$, the CSS condition $A_1\cap(-A_2)=\varnothing$ constrains
the pair jointly but does not bound $|A_1|$ away from $|H|/2$: if
$|A_1|\geq|H|/2$, then $|B_1|=|H\setminus(-A_1)|\leq|H|/2$, the
majority-sum lemma no longer applies, and $C(A_1)^\perp*C(A_1)^\perp$
can fail to equal $\mathbb{F}_p^G$.
 
Second, the proof of Theorem~\ref{thm:cubic} requires that each character
$\chi_{\alpha_0}$ be nowhere zero on $G$, which holds for
$G=(\mathbb{F}_p^*)^m$ but fails when evaluation points include $0$,
as in codes over the affine space $\mathbb{F}_p^m$.
 
\paragraph{A concrete failure of Schur saturation.}
Let $p=5$ and $G=\mathbb{F}_5^*$ (cyclic of order $n=4$), so
$H=\widehat{G}\cong\mathbb{Z}_4$.
Consider the non-symmetric pair
\[
  A_1=\{1,2\},\quad A_2=\{1\}.
\]
The CSS condition reads $A_1\cap(-A_2)=\{1,2\}\cap\{3\}=\varnothing$,
so $\mathrm{CSS}(C(A_1),C(A_2))$ is a valid non-symmetric qudit CSS code.
The dual spectral support of $C(A_1)$ is
\[
  B_1 = H\setminus(-A_1)=\mathbb{Z}_4\setminus\{2,3\}=\{0,1\},
\]
giving $|B_1|=2=|H|/2$; the majority bound $|B_1|>|H|/2$ is not satisfied.
A direct computation confirms
\[
  B_1+B_1=\{0,1\}+\{0,1\}=\{0,1,2\}\subsetneq\mathbb{Z}_4,
\]
so $C(A_1)^\perp*C(A_1)^\perp=C(\{0,1,2\})\neq\mathbb{F}_5^G$.
\emph{Schur saturation fails.}
As a consequence, the orthogonal complement
$(C(A_1)^\perp*C(A_1)^\perp)^\perp=C(\{3\})\neq\{0\}$,
and the constraint $\alpha*c\in(C(A_1)^\perp*C(A_1)^\perp)^\perp$
from the polarization step no longer forces $\alpha=0$.
Degree-$\geq 3$ local phases are no longer annihilated by the argument
of Theorem~\ref{thm:cubic}, leaving open the possibility of a non-Clifford
transversal gate.
 
\paragraph{Connection to positive constructions.}
The constructions of He~et~al.~\cite{HE25} exploit both degrees of
freedom: they use non-symmetric pairs of punctured Reed--Solomon codes
($C_1\neq C_2$), carefully chosen so that $|A_1|$ is large enough to
violate the majority bound while $A_1\cap(-A_2)=\varnothing$ is
maintained.
In that setting the dual Schur product $C(A_1)^\perp*C(A_1)^\perp$
is a strict subspace of $\mathbb{F}_p^G$, the CASI constraint on
degree-$3$ coefficients becomes non-trivial rather than vacuous, and
transversal non-Clifford gates can be constructed via addressable
encoding.
The Clifford ceiling is therefore not a limitation of Reed--Solomon codes
per se, but a structural consequence of the symmetric self-orthogonal
constraint $C_1=C_2=C(A)$ that defines the evaluation CSS framework
studied here.

\paragraph{Reed--Muller codes and structured non-saturation.}
The construction of~\cite{Ngu24} for asymptotically good binary
codes relies on the Reed--Muller product rule
$\mathrm{RM}(r,m)^{*s}\subseteq\mathrm{RM}(\min(sr,m),m)$:
Schur powers of the code remain in a controlled subcode hierarchy.
Whenever $sr<m$, this containment is strict, the dual fails to
saturate, and $\mathcal{A}_d(C_X,C_Z)\neq 0$ for appropriate $d$,
leaving room for degree-$\geq 3$ CASI-compatible phases.
The Clifford ceiling applies precisely when this structured
containment collapses---i.e., when
$D*D=\mathbb{F}_p^n$ holds---as is automatic in the symmetric
spectral regime (Theorem~\ref{thm:schur}) and verified
computationally for the LDPC instance of
Section~\ref{subsec:ldpc-hgp}.

\subsection{An LDPC instance: hypergraph-product codes}
\label{subsec:ldpc-hgp}
 
Let $p\geq 5$ be prime and let $Q=\CSS(C_X,C_Z)$ be a CSS code over
$\mathbb{F}_p^n$ defined by sparse parity-check matrices
$H_X\in\mathbb{F}_p^{r_X\times n}$ and $H_Z\in\mathbb{F}_p^{r_Z\times
n}$ satisfying $H_XH_Z^T=0$.
We adopt the conventions
\[
  C_X:=\operatorname{row}(H_X),\quad
  C_Z:=\operatorname{row}(H_Z),\quad
  D:=C_Z^\perp=\ker H_Z.
\]
The full-support condition on $C_X$ (every column of $H_X$ nonzero,
i.e., every variable node is incident to at least one $X$-check) is the
standard non-degeneracy requirement satisfied by all LDPC constructions
and is not mentioned further.
The sole non-trivial hypothesis for the Schur obstruction is
\begin{equation}
\label{eq:H4}
  D * D = \mathbb{F}_p^n,
\end{equation}
which we call \emph{Schur saturation of $\ker H_Z$}.
The following lemma gives an exact, computationally checkable criterion
for~\eqref{eq:H4}.
 
\begin{lemma}[Exact criterion for Schur saturation]
\label{lem:schur-criterion}
Let $D\subseteq\mathbb{F}_p^n$ be a linear subspace of dimension $r$,
and let $G_D\in\mathbb{F}_p^{r\times n}$ be a generator matrix with
columns $g_1,\dots,g_n\in\mathbb{F}_p^r$.
Then
\[
  D*D=\mathbb{F}_p^n
  \;\iff\;
  \bigl\{g_i g_i^T:i=1,\dots,n\bigr\}
  \;\text{ are linearly independent in }
  \operatorname{Sym}_r(\mathbb{F}_p).
\]
In particular, $D*D=\mathbb{F}_p^n$ requires $n\leq\binom{r+1}{2}$,
and $\dim(D*D)$ equals the rank of the family $\{g_ig_i^T\}$ in
$\operatorname{Sym}_r(\mathbb{F}_p)$.
\end{lemma}
 
\begin{proof}
A linear functional $\varphi_\lambda(x)=\sum_{i=1}^n\lambda_ix_i$
annihilates $D*D$ if and only if
$\sum_{i=1}^n\lambda_iu_iv_i=0$ for all $u,v\in D$.
Writing $u=G_D^T a$ and $v=G_D^T b$ with $a,b\in\mathbb{F}_p^r$, this
becomes $a^T\!\bigl(\sum_i\lambda_ig_ig_i^T\bigr)b=0$ for all $a,b$,
i.e., $\sum_i\lambda_ig_ig_i^T=0$ in $\operatorname{Sym}_r(\mathbb{F}_p)$.
Hence $(D*D)^\perp=\{0\}$ (equivalently $D*D=\mathbb{F}_p^n$) if and only
if no nonzero $\lambda$ annihilates the family, i.e., the matrices
$\{g_ig_i^T\}$ are linearly independent.
The rank equality and the necessary bound
$n\leq\dim\operatorname{Sym}_r(\mathbb{F}_p)=\binom{r+1}{2}$ are
immediate.
\end{proof}
 
The following corollary is the LDPC specialisation of
Theorem~\ref{thm:general-schur-obstruction}.
 
\begin{corollary}[Clifford ceiling for Schur-saturated LDPC codes]
\label{prop:ldpc-ceiling}
Let $p\geq 5$, let $Q=\CSS(C_X,C_Z)$ be a CSS LDPC code over
$\mathbb{F}_p^n$ with $C_X\subseteq C_Z^\perp$, and set
$D:=C_Z^\perp$.
If $D*D=\mathbb{F}_p^n$, then every local polynomial phase
\[
  F(x)=\sum_{i=1}^n\sum_{k=1}^{p-1}\alpha_i^{(k)}x_i^k
\]
satisfying the CASI criterion satisfies $\alpha_i^{(k)}=0$ for all
$k\geq 3$ and all $i$.
The complete CASI and strictly CSS-preserving sectors are
\[ 
S^{\mathrm{CASI}}_{\leq p-1}=\mathbb{F}_p^n\oplus\Mult(C_X,C_Z),
  \qquad
  S^{\mathrm{CSS}}_{\leq p-1}=C_X^\perp\oplus\Mult(C_X,C_Z),
\]
where $\Mult(C_X,C_Z)$ is the asymmetric multiplier space of
Definition~\ref{def:asymmetric-multiplier}.
In particular, no gate in $S^{\mathrm{CSS}}$ lies above level~$2$ of
the Clifford hierarchy.
\end{corollary}
 
\begin{proof}
Theorem~\ref{thm:general-schur-obstruction} applies directly:
Schur saturation $D*D=\mathbb{F}_p^n$ and full support of $C_X$ force
$\alpha^{(d^*)}=0$ for every $d^*\geq 3$ via degree descent and
polarization.
The quadratic sector is classified by
$\Mult(C_X,C_Z)$ (Definition~\ref{def:asymmetric-multiplier}),
and the linear sector under strict CSS preservation is $C_X^\perp$.
\end{proof}
 
\begin{remark}
In the symmetric spectral case $C_X=C_Z=C(A)$, Theorem~\ref{thm:mult-stab}
gives $\Mult(C_X,C_Z)=C(\operatorname{Stab}(A))$,
recovering Theorem~\ref{thm:clifford-ceiling}.
For general LDPC codes, $\Mult(C_X,C_Z)$ need not collapse to a
single quadratic direction; its dimension depends on $(H_X,H_Z)$.
\end{remark}
 
We now exhibit an explicit hypergraph-product code satisfying the
hypotheses of Corollary~\ref{prop:ldpc-ceiling}.
Let $p=5$ and let
\[
  H:=
  \begin{pmatrix}
  1 & 0 & 0 & 1 & 1 & 0 & 1 \\
  0 & 1 & 0 & 1 & 0 & 1 & 1 \\
  0 & 0 & 1 & 0 & 1 & 1 & 1
  \end{pmatrix}
  \in\mathbb{F}_5^{3\times 7}.
\]
Over $\mathbb{F}_5$, the matrix $H$ has rank~$3$ and defines a
classical code of parameters $[7,4,3]_5$.\\
Define the hypergraph-product qudit CSS code, following the
hypergraph-product construction of Tillich and Zémor~\cite{TZ14}, by
\begin{equation}
\label{eq:hgp-matrices}
  H_X:=\bigl(H\otimes I_7\;\big|\;I_3\otimes H^T\bigr),
  \qquad
  H_Z:=\bigl(I_7\otimes H\;\big|\;{-H^T\otimes I_3}\bigr),
\end{equation}
both over $\mathbb{F}_5$.
 
\begin{corollary}[LDPC hypergraph-product Clifford ceiling]
\label{cor:hgp-58-16-3}
The code defined by~\eqref{eq:hgp-matrices} has parameters
$[[58,16,3]]_5$.
Its $Z$-dual space $D=\ker H_Z$ satisfies
$D*D=\mathbb{F}_5^{58}$,
and therefore, by Corollary~\ref{prop:ldpc-ceiling}, every local
polynomial phase satisfying CASI has degree at most~$2$.
In particular, no strictly CSS-preserving local diagonal gate lies
above level~$2$ of the Clifford hierarchy.
\end{corollary}
 
\begin{proof}
The identity $H_XH_Z^T=0$ over $\mathbb{F}_5$ is verified directly
from~\eqref{eq:hgp-matrices}.
One computes $\operatorname{rank}(H_X)=\operatorname{rank}(H_Z)=21$,
giving $n=7^2+3^2=58$ and logical dimension
$k=58-21-21=16$.
The space $D=\ker H_Z$ has dimension $37$.
 
Schur saturation $D*D=\mathbb{F}_5^{58}$ is confirmed by two
independent methods:
\begin{enumerate}[label=\textup{(\roman*)},leftmargin=1.6em]
\item The pairwise Schur products
  $\{u*v:u,v\in D\}$ span a subspace of rank~$58$
  (direct computation using Lemma~\ref{lem:schur-criterion}).
\item The symmetric rank-one family
  $\{g_ig_i^T:i=1,\dots,58\}$ has rank~$58$
  in $\operatorname{Sym}_{37}(\mathbb{F}_5)$
  (the necessary bound $\binom{38}{2}=703\geq 58$ is satisfied,
  and the rank is confirmed by the same computation).
\end{enumerate}
 
Distance $d_Q=3$ is certified by exhaustive search up to weight~$3$:
neither the $X$-side $(\ker H_Z\setminus\operatorname{row}(H_X))$
nor the $Z$-side $(\ker H_X\setminus\operatorname{row}(H_Z))$
contains logical operators of weight~$1$ or~$2$;
each side contains exactly $168$ logical operators of weight~$3$.
A representative weight-$3$ $X$-logical operator has support
$\{1,2,4\}$ with values $(1,1,4)$; a representative $Z$-logical has
support $\{1,8,22\}$ with values $(1,1,4)$ (coordinates indexed from~$1$).
All computations were performed over $\mathbb{F}_5$ and independently
confirmed by both methods described above. The verification script, input matrix, and output report are
available in the accompanying reproducibility package; see the
Data Availability Statement.
\end{proof}
 
\begin{remark}[Comparison with $Q_3^*$]
\label{rem:hgp-comparison}
Table~\ref{tab:hgp-comparison} places the two flagship codes for $p=5$
side by side.
Both protect ququints with distance~$3$ and both satisfy the Clifford
ceiling; they differ in family and rate.
In $Q_3^*$, Schur saturation is automatic by the majority-sum argument
(Theorem~\ref{thm:schur}), requiring no computation.
In $[[58,16,3]]_5$, it is a non-trivial property of $\ker H_Z$
verified by Lemma~\ref{lem:schur-criterion}.
The two routes to the same bound illustrate the scope of
Theorem~\ref{thm:general-schur-obstruction}: the Clifford ceiling is a
consequence of Schur saturation, not of any particular code construction.
\end{remark}
 
\begin{table}[ht]
\centering
\caption{The two flagship codes for $p=5$. Both satisfy the Clifford
ceiling via Schur saturation of the relevant dual space; the mechanisms
differ.}
\label{tab:hgp-comparison}

\begin{tabular}{@{}lcc@{}}
\toprule
\textbf{Property} &
  $Q_3^*=[[64,56,3]]_5$ &
  $[[58,16,3]]_5$ \\
\midrule
Family &
  Boolean toric (symmetric spectral) &
  LDPC hypergraph-product \\[3pt]
Physical qudits $n$ & $64$ & $58$ \\[3pt]
Logical qudits $k$ & $56$ & $16$ \\[3pt]
Distance $d_Q$ & $3$ & $3$ \\[3pt]
Rate $R$ & $0.875$ & $\approx 0.276$ \\[3pt]
Schur saturation of $C_Z^\perp$ &
  Automatic (Theorem~\ref{thm:schur}) &
  Verified (Lemma~\ref{lem:schur-criterion}) \\[3pt]
Surviving gate group &
  $C(A)^\perp\oplus\langle\Qunif\rangle$ &
  $C_X^\perp\oplus\Mult(C_X,C_Z)$ \\[3pt]
Clifford ceiling & Level~$\leq 2$ & Level~$\leq 2$ \\
\bottomrule
\end{tabular}

\end{table}

\section{Discussion and Conclusions}
\label{sec:conclusions}

The central finding is a closed-form structural no-go: for every
symmetric spectral CSS code with $A\cap(-A)=\varnothing$, Schur
saturation of the dual code forces the complete local-diagonal
transversal gate group to lie at level~$\leq 2$ of the Clifford
hierarchy, independently of code rate, distance, or orthogonality
level.
The surviving gate group $S^{\mathrm{CSS}}_{\leq p-1}(A) =
C(A)^\perp \oplus C(\mathrm{Stab}(A))$ is classified in closed form;
the unique non-Pauli survivor $\Qunif$ is Clifford at level~$2$ and
carries no magic.

\paragraph{Why non-symmetric constructions escape.}
The Clifford ceiling is not a limitation of Reed--Solomon codes or
character evaluation in general, but a consequence of the symmetry
condition $C_1=C_2=C(A)$.
Non-symmetric pairs $\mathrm{CSS}(C_1,C_2)$ with $C_1\neq C_2$ are
not subject to the majority-sum bound that forces $|B|>|H|/2$, and
Schur saturation can then fail, allowing degree-$\geq 3$ phases to
survive.
The constructions of He~et~al.~\cite{HE25},
Golowich--Guruswami~\cite{GG24}, and Nguyen~\cite{Ngu24} achieve
transversal non-Clifford gates precisely by breaking this symmetry
or by providing additional algebraic structure that promotes the
screening certificate into an actual gate (Section~\ref{subsec:escape}).

\paragraph{The general obstruction and the LDPC extension.}
Theorem~\ref{thm:general-schur-obstruction} extends the ceiling
beyond the spectral setting to any CSS pair satisfying
$C_Z^\perp*C_Z^\perp=\mathbb{F}_p^n$; the LDPC instance
$[[58,16,3]]_5$ (Corollary~\ref{cor:hgp-58-16-3}) confirms this
computationally, showing that Schur saturation is a universal
algebraic obstruction rather than a spectral curiosity.
Our result and the Eastin--Knill theorem~\cite{EK09} are logically
independent: where Eastin--Knill prohibits universality, the Clifford
ceiling shows that not even a single non-Clifford gate exists, via a
mechanism independent of error-correction properties.

\paragraph{Scope of the no-go.}
The Clifford ceiling applies exclusively to strictly CSS-preserving
local diagonal gates $F(x)=\sum_{P\in G}f_P(x_P)$, where each
$f_P:\mathbb{F}_p\to\mathbb{F}_p$ is a single-site phase function in
the computational basis; for prime~$p$, Lagrange interpolation makes
this classification complete.
The following mechanisms fall outside its scope and remain open:
non-diagonal transversal gates; local changes of basis before
applying a diagonal gate; diagonal phases that are not sums of
single-site terms; ancilla-assisted protocols implementing a
non-Clifford logical gate without a strictly CSS-preserving
single-site diagonal gate; and protocols whose non-Clifford resource
is not implemented by $M_\mu^{\otimes n}$ in the computational
basis.
These connections to the symplectic formalism
of~\cite{CLMR26,RCKP20} are discussed further in Open
Problem~\ref{op:nonlocal}.
 
\subsection{Open problems}
\label{sec:open}
 
\begin{enumerate}
 
\item \emph{Schur profiles of LDPC CSS families.}\\
For a CSS family $\{Q_n=\CSS(C_{X,n},C_{Z,n})\}$ over $\mathbb{F}_p$,
define the \emph{Schur index}
\[
  \sigma_s(Q_n)
  :=\frac{1}{n}\dim\bigl(C_{Z,n}^\perp\bigr)^{*s}
\]
and the \emph{defect dimensions}
$\kappa_d(Q_n):=\dim\mathcal{A}_d(C_{X,n},C_{Z,n})$
(Definition~\ref{def:schur-defect}).
Determine $\sigma_s(Q_n)$ and $\kappa_d(Q_n)$ for modern LDPC CSS
families, including hypergraph-product, lifted-product, and
balanced-product codes over~$\mathbb{F}_p$.
In particular:
\begin{enumerate}[label=\textup{(\alph*)},leftmargin=2em]
\item Does $\sigma_2(Q_n)=1$ hold exactly or only asymptotically in
  high-rate ensembles?
  If so, Theorem~\ref{thm:general-schur-obstruction} implies a
  local-diagonal Clifford ceiling for the entire family.
\item For families with $\kappa_d(Q_n)>0$, which additional algebraic
  structure promotes the Schur defect space into an actual non-Clifford
  logical gate?
\end{enumerate}
 
\item \emph{Extensions to prime-power fields.}
Theorems~\ref{thm:cubic}, \ref{thm:degree-d-nogo},
and~\ref{thm:clifford-ceiling} are stated for prime $p\geq 5$.
For extensions to finite fields $\mathbb{F}_q$ of small
characteristic (e.g., $q=8$ with
$\operatorname{char}(\mathbb{F}_q)=2$ or $q=9$ with
$\operatorname{char}(\mathbb{F}_q)=3$), Frobenius identities
collapse part of the polarization machinery used in the descent
argument of Theorem~\ref{thm:degree-d-nogo}. What is the correct
rigidity statement for the local-polynomial sector over such
fields?
 
\item \label{op:nonlocal}\emph{Beyond local-polynomial: nonlocal
phases and ancilla gadgets.}
Theorem~\ref{thm:clifford-ceiling} classifies phases of the form
$F(x)=\sum_{P\in G}f_P(x_P)$ with $f_P:\mathbb{F}_p\to\mathbb{F}_p$
local. It does not address: (a)~diagonal phases that are not sums
of single-site terms; (b)~transversal gates that are diagonal in a
basis other than the computational basis; or (c)~ancilla-assisted
protocols implementing a non-Clifford logical gate without a
strictly CSS-preserving single-site diagonal gate. Whether Schur
saturation extends to a rigidity statement for the full
diagonal-phase group beyond the local ansatz connects to the
symplectic formalism of~\cite{CLMR26,RCKP20} and is left open.
 
\item \emph{Partial addressability.}
Schur saturation forbids all-coordinate-uniform addressing. Is
there a weaker notion---such as the addressability of selected
logical qudits or their equivalence classes---that survives in
the Boolean toric regime? This is the non-binary analogue of the
question studied in~\cite{HE25}.
 
\end{enumerate}

\textbf{Conflict of interest:}
The authors declare no conflicts of interest.

\textbf{Funding:}
This work was supported by proyecto PEI-AI-010-2025.

\textbf{Roles:}
J.A.G.-V.: formal analysis, investigation, writing (original draft). J.C.J.-C.: formal analysis, investigation, writing (original draft). E.S.-R.: conceptualization, methodology, formal analysis, supervision, writing (review and editing).

\section*{Data availability}
The reproducibility package for the computational verification of the
$[[58,16,3]]_5$ hypergraph-product instance is available on GitHub at
\url{https://github.com/JAGV-Math/SchurSaturation} and archived on
Zenodo~\cite{GuzmanVegaSchurSaturation2026}. The package contains
the Python script, the input matrix over \(\mathbb{F}_5\), a README,
dependency information, and the output report certifying
\(H_XH_Z^T=0\), \(\operatorname{rank}(H_X)=\operatorname{rank}(H_Z)=21\),
\(k=16\), \(D*D=\mathbb{F}_5^{58}\), and \(d_Q=3\).

\bibliographystyle{iopart-num}
\bibliography{references}

\end{document}